\documentclass[aps,twocolumn,superscriptaddress,showpacs,showkeys,preprintnumbers,amsmath,amssymb,amsfonts,nofootinbib]{revtex4}

\pdfoutput=1

\usepackage{subfigure}
\usepackage{graphicx}
\usepackage{color}
\usepackage{slashed}
\usepackage{tabularx}
\usepackage{hyperref}
\newcommand{\Tr}{\mathrm{Tr}}

\newcommand{\I}{\mathrm{i}}

\newcolumntype{L}[1]{>{\raggedright\arraybackslash}p{#1}} 
\newcolumntype{C}[1]{>{\centering\arraybackslash}p{#1}} 
\newcolumntype{R}[1]{>{\raggedleft\arraybackslash}p{#1}} 
\graphicspath{{.}{figures/}{mma/}}

\definecolor{blue}{rgb}{0,0,1}

\definecolor{green}{rgb}{0,1,0}

\definecolor{red}{rgb}{1,0,0}

\def\Eq#1{Eq.~(\ref{#1})}
\def\Sec#1{Sec.~\ref{#1}}
\def\Fig#1{Fig.~\ref{#1}}
\def\App#1{App.~\ref{#1}}

\begin{document}

\title{In-Medium Spectral Functions of Vector- and Axial-Vector Mesons from the Functional Renormalization Group}

\newcommand{\TUD}{Theoriezentrum, Institut f\"ur Kernphysik, Technische Universit\"at Darmstadt, Schlossgartenstrasse 2, 64289 Darmstadt, Germany}
\newcommand{\JLU}{Institut f\"ur Theoretische Physik, Justus-Liebig-Universit\"at Giessen, Heinrich-Buff-Ring 16, 35392 Giessen, Germany}
\newcommand{\GSI}{GSI Helmholtzzentrum f\"ur Schwerionenforschung GmbH, Planckstraße 1, 64291 Darmstadt, Germany}
\newcommand{\ECT}{European Centre for Theoretical Studies in Nuclear Physics and related Areas (ECT*) and Fondazione Bruno Kessler, Villa Tambosi, Strade delle Tabarelle 286, I-38123 Villazzano (TN), Italy}
\newcommand{\ITPHD}{Institut f\"ur Theoretische Physik, Ruprecht-Karls-Universit\"at Heidelberg, Philosophenweg 16, 69120 Heidelberg, Germany}

\author{Christopher Jung}\affiliation{\JLU}\affiliation{\TUD}
\author{Fabian Rennecke}\affiliation{\JLU}\affiliation{\ITPHD}
\author{Ralf-Arno Tripolt}\affiliation{\ECT}
\author{Lorenz von Smekal}\affiliation{\JLU}
\author{Jochen Wambach}\affiliation{\TUD}\affiliation{\ECT}

\begin{abstract}
In this work we present first results on vector and axial-vector meson spectral functions as obtained by applying the non-perturbative functional renormalization group approach to an effective low-energy theory motivated by the gauged linear sigma model. By using a recently proposed analytic continuation method, we study the in-medium behavior of the spectral functions of the $\rho$ and $a_1$ mesons in different regimes of the phase diagram. In particular, we demonstrate explicitly how these spectral functions degenerate at high temperatures as well as at large chemical potentials, as a consequence of the restoration of chiral symmetry. In addition, we also compute the momentum dependence of the $\rho$ and $a_1$ spectral functions and discuss the various time-like and space-like processes that can occur.
\end{abstract}



\maketitle

\section{Introduction}\label{sec:intro}

A major challenge in Quantum Chromodynamics (QCD) is to explore the phase structure of strong-interaction matter, including possible phase transitions and the existence and location of a critical endpoint (CEP) \cite{Stephanov:2004wx,Braun-Munzinger2009}. In experiments one performs heavy-ion collisions to produce extremely high energy densities leading to new phases such as the quark-gluon plasma (QGP). To get insights into the entire space-time history of the collision process, real photons and dileptons are particularly useful probes, since they have negligible interaction with the hadronic fireball \cite{Rapp:2013nxa,vanHees:2009vk,Rapp:2013ema}. In this context the decay of vector mesons, located in the low invariant-mass regime, and here especially the $\rho$ meson, is interesting, since the quantum numbers of vector mesons allow them to directly decay into dileptons \cite{RappWambach1999}. For this reason the in-medium properties  of the $\rho$ meson have received considerable attention \cite{Brown1991,Hohler:2013eba,Rapp:1997fs}. By analyzing dilepton spectra at low invariant masses one tries to find evidence for chiral symmetry restoration or the existence of a CEP \cite{Pisarski:1981mq, Rapp:2009yu}. 

The basis for calculating vector-meson spectral functions is a physically reasonable description of vector mesons within effective models for QCD. The first model including vector mesons was introduced by Sakurai in the 1960's \cite{1960AnPhy} who generalized the known gauge principle from QED to local $SU(2)_V$ isospin symmetry present in QCD. This concept was in accordance with experimental observations and is referred to as the vector meson dominance model (VMD) \cite{Schildknecht:2005xr}.
To study chiral symmetry and its breaking pattern in more detail, the gauge concept of Sakurai was extended to the chiral symmetry group $SU(2)_L\times SU(2)_R$ as a local gauge symmetry, where the $a_1$ meson as chiral partner of the $\rho$ meson also appears as a gauge boson \cite{Lee:1967ug} thus leading to the `gauged linear sigma model'. Another possibility is to impose a global chiral symmetry rather than a local one \cite{Urban:2001ru, Rennecke:2015eba, Eser:2015pka}.

The feasibility of calculating in-medium spectral functions from the functional renormalization group (FRG)  \cite{Polonyi:2001se,Bagnuls:2000ae,Gies:2006wv,Berges:2000ew,Pawlowski:2005xe,Litim:2006ag,vonSmekal:2012vx,Pawlowski:2010ht,Delamotte:2007pf} has been demonstrated in \cite{Tripolt2014,Tripolt2014a,TripoltSmekalWambach2016a}. Apart from the absence of a fermion sign problem at finite chemical potential, one of the main advantages of the method proposed is that the analytic continuation from Euclidean to Minkowski space-time can be performed in a well-defined and simple way. Moreover, since thermal and quantum fluctuations are taken into account properly within the FRG approach, the method is also well suited to treat critical phenomena like phase transitions \cite{Schaefer:2004en,Schaefer:2006sr,Schaefer:2007pw,Herbst:2010rf,Strodthoff:2011tz, Mitter:2013fxa, Pawlowski2014, Khan:2015puu}. Aside  from the non-perturbative method, presented here, there are 
also intriguing phenomenological approaches to address the degeneracy of vector- and axialvector spectral functions based on sum rules and loop expansions of gauged chiral Lagrangians \cite{Holt:2012wr,Hohler:2015iba}.

In this paper we present first results for the $\rho$ and $a_1$ meson spectral functions from the FRG based on the gauged linear sigma model, inspired by \cite{Rennecke:2015eba}. We derive flow equations for the real-time two-point functions of $\rho$ and $a_1$ mesons which are obtained from their Euclidean counterparts via an analytic continuation on the level of the flow equations \cite{Floerchinger2012, Tripolt2014,Kamikado2014, Pawlowski:2015mia}. This allows access to real-time quantities such as pole masses and  decay widths which we calculate in various regions of the phase diagram predicted by the model, especially along the axis of vanishing chemical potential and  the $\mu$-axis across the CEP. Since our FRG treatment is thermodynamically consistent and symmetry preserving, the in-medium modifications of the spectral functions can be stringently connected to the restoration of chiral symmetry, because the order parameter for chiral symmetry breaking and the spectral functions are obtained self-consistently. 

The paper is organized as follows. We first motivate the model used and our ansatz for the quantum effective action in \Sec{subsec:model}. We then discuss the salient features of the FRG as a non-perturbative method as well as the resulting flow equations in \Sec{subsec:frg}. After going through some details concerning the numerical implementation in \Sec{subsec:numerics} we discuss the phase structure of the model as well as the $T$- and $\mu$-dependent Euclidean and pole masses of the particles in \Sec{subsec:masses}. We then present $\rho$ and $a_1$ spectral functions at finite temperature and chemical potential, first for vanishing external spatial momentum in \Sec{subsec:spectral_function} and then for finite external spatial momentum in \Sec{subsec:results_p}. We conclude by giving a summary in \Sec{sec:summary} and present more details concerning the derivation of the flow equations, the analytic continuation procedure, the available processes and explicit expressions in Apps.~\ref{sec:flow_equations}-\ref{sec:loop_functions}.

\section{Theoretical setup}\label{sec:setup}

\subsection{Extended linear-sigma model with quarks}\label{subsec:model}

In this section we will present a simple low-energy model of two-flavor QCD that captures the main features relevant for the description of the $\rho$ and $a_1$ meson which is based on an previous work concerning the description of vector mesons in QCD \cite{Rennecke:2015eba}. In that work the connection of the effective action with QCD is given by successively integrating out quantum fluctuations starting from QCD at high energies.  We focus on the dynamical generation of mesons but note that a similar reasoning can be applied to the formation of baryons.

The increasing strength of the strong coupling $\alpha_s$ with decreasing energy scale leads to strong two-quark--two-antiquark correlations. This naturally gives rise to effective quark-antiquark scattering channels $\lambda_i (\bar\psi T_i \psi)^2$ with the effective couplings $\lambda_i \!\propto\! \alpha_s^2$ and the tensor structures $T_i$. Hence, the QCD effective action naturally assumes the form of a gauged Nambu--Jona-Lasinio (NJL) model. With increasing $\alpha_s$, also the $\lambda_i$ increase until they eventually diverge. These divergences signal the formation of bound states and resonances in the quark-antiquark scattering channels. Since this applies in particular to the scalar-isoscalar channel, chiral symmetry breaking is dynamically generated.

We compute spectral functions from FRG flows which, in essence, are based on the fluctuations of off-shell degrees of freedom, see Sec.~\ref{subsec:frg}. Hence, the lightest particles give the most relevant contributions in the low-energy regime. For two quark flavors the dynamically most relevant mesons are pions as they are the pseudo-Goldstone bosons of spontaneous chiral symmetry breaking. Their chiral partner, the scalar $\sigma$-meson, has to be included into our effective description as well. For two flavors, chiral $SU(2)_L\times SU(2)_R$ symmetry is locally isomorphic to $O(4)$ and we need both the isotriplet of pions, $\vec{\pi}$, and the isoscalar $\sigma$ for the construction of chiral invariants. Furthermore, the $\sigma$ mode is the critical mode at the CEP, i.e~it becomes exactly massless there. To capture chiral symmetry restoration in the spectral functions of the phenomenologically relevant $\rho$ meson, we also include its chiral partner the $a_1$ meson. The corresponding four-quark channels of the effective NJL-type action are given by
\begin{align}\label{eq:4quark}
\begin{split}
\mathcal{L}_{(4q)} &= \frac{\lambda_S}{2} \Big[\big(\bar\psi \psi\big)^2 - \big(\bar\psi \gamma_5 \vec{\tau}\psi\big)^2\Big]\\
&\quad- \frac{\lambda_V}{2} \Big[\big(\bar\psi \gamma_\mu \vec{\tau} \psi\big)^2 - \big(\bar\psi \gamma_\mu \gamma_5 \vec{\tau}\psi\big)^2\Big]\,,
\end{split}
\end{align}
where $\vec{\tau}$ are the Pauli matrices.
The first term of \Eq{eq:4quark} has the quantum numbers of $\sigma$-
and $\pi$-meson and the second term that of $\rho$ and $a_1$-meson,
respectively. The formation of mesons at the scale
of chiral symmetry breaking will be reflected by poles in the
corresponding quark-antiquark scattering channel.
Their properties at low energies can in principle be studied by computing the full momentum dependence of $\lambda_S$ and $\lambda_V$. We resort to a simpler treatment by explicitly introducing mesons as they are the dominant low-energy degrees of freedom. This is done by partially bosonizing the four-quark interaction \Eq{eq:4quark} by means of a Hubbard-Stratonovich transformation. This yields the following mixed fermionic-bosonic contribution to the effective action,
\begin{align}\label{eq:fb}
\begin{split}
\mathcal{L}_{(\text{FB})} &= \frac{1}{2} m_S^2\big(\sigma^2+\vec{\pi}^{\,2}\big) +  h_S \bar\psi\big(\sigma + i \gamma_5\vec{\tau}\vec{\pi} \big)\psi\\
&\quad + \frac{1}{2} m_V^2 \big[(\vec{\rho}^{\,\mu})^2+(\vec{a}_1^{\,\mu})^2\big]\\
&\quad + i h_V \bar\psi\big(\gamma_\mu\vec{\tau}\vec{\rho}^{\,\mu} + \gamma_\mu\gamma_5\vec{\tau}\vec{a}_1^{\,\mu} \big)\psi\,,
\end{split}
\end{align}
which directly reflects the fermionic pairing through the scalar- and vector Yukawa couplings $h_S$ and $h_V$. In the present case, the scale of bosonization is chosen to be the UV-cutoff $\Lambda$ of our effective description. There, \Eq{eq:4quark} is equivalent to \Eq{eq:fb} also on the quantum level if
\begin{align}\label{eq:hlm}
\lambda_S = \frac{h_S^2}{m_S^2}\,,\quad\text{and}\quad \lambda_V = \frac{h_V^2}{m_V^2}\,.
\end{align}
Note that chiral symmetry breaking, i.e.~a diverging $\lambda_{S}$, is
signaled by a sign change of the mass parameters $m^2_{S}$ in analogy
to Ginzburg-Landau theory. In QCD, the scale of meson formation
emerges dynamically and is not introduced by hand. To capture this
transition within a unified framework, one can use dynamical
hadronization as put forward for QCD in \cite{Braun:2014ata,Mitter:2014wpa}
and applied to vector mesons in the vacuum in \cite{Rennecke:2015eba}. Since we are only interested in the low-energy effective theory here, we postpone a treatment within full QCD to future work.

If we further integrate out fluctuations, the mesons which were formally introduced as auxiliary fields in \Eq{eq:fb} become dynamical and higher-order effective meson interactions are generated. Furthermore, the gauge sector of QCD develops a mass gap and eventually decouples from the system at low energies. Here we assume that the gluons are fully integrated out at the UV-cutoff $\Lambda$ of the low-energy effective theory.

While, in principle, all meson interactions consistent with the global and local symmetries of the system are present in the mesonic part of the effective action, we restrict ourselves to the convenient case of a gauged linear sigma model \cite{Lee:1967ug}. To this end, we assume that the meson sector has local a chiral $SU(2)_L\times SU(2)_R$ symmetry. This assumption is particularly powerful since the interactions involving vector mesons are completely fixed by gauge symmetry. This reduces the potentially large number of effective couplings to one gauge coupling $g$. Putting all this together finally leads us to the following effective action:
\begin{widetext}
\begin{align}
\label{eq:effective_action}
\Gamma_k = \int d^4x &\Big[\bar{\psi} \left(\slashed{\partial}-\mu\gamma_0+
h_S\left(\sigma +\mathrm{i} \vec{\tau}\vec{\pi}\gamma_5\right)+
\mathrm{i} h_V \left(\gamma_{\mu} \vec{\tau}\vec{\rho}^{\mu}+\gamma_{\mu}\gamma_5 \vec{\tau}\vec{a}_1^{\mu}\right)
\right)\psi+ U_{k}(\phi^2)-c\sigma+\frac{1}{2} (\partial_{\mu}\phi)^2 \nonumber\\
&+ \frac{1}{8} \text{Tr}\left(\partial_{\mu}V_{\nu}-\partial_{\nu}V_{\mu}\right)^2-
\mathrm{i}g V_{\mu}\phi\partial_{\mu}\phi-\frac{1}{2}g^2\left(V_{\mu}\phi\right)^2+\frac{1}{4}m_{k,V}^2 \text{Tr}\left(V_{\mu}V_{\mu}\right)
\Big] +\Delta\Gamma_{\pi a_1}\,,
\end{align}
\end{widetext}
where $U_{k}(\phi^2)$ is the $O(4)$ symmetric effective potential and
a function of the chiral invariant $\phi^2$ with
$\phi=(\vec{\pi},\sigma)^T$. This constitutes the lowest order in a
gradient expansion of the effective action, and implies in particular
that wavefunction renormalizations are not taken into account. The
source term $-c\sigma$ stems from the bosonization of the
current-quark mass term 
of the QCD action. It therefore accounts for the explicit chiral
symmetry breaking through explicit quark masses. Hence, the pions are
pseudo-Goldstone bosons with finite mass and the chiral phase
transition is a crossover at small chemical potentials. 

The vector mesons are given in the adjoint representation of $O(4)$ with
\begin{align}
V_\mu = \vec{\rho}^{\,\mu}\vec{T}+\vec{a}_1^{\,\mu}\vec{T}^5\,.
\end{align}
We define the $\mathfrak{so}(4)$ matrices
\begin{align}
(T_i)_{jk}=\begin{pmatrix} -i\epsilon_{ijk} & \vec{0} \\ \vec{0}^T & 0\end{pmatrix},\quad (T_i^5)=\begin{pmatrix} 0_{3 \!\times\! 3} & -i\vec{e}_i \\ i\vec{e}_i^{\,T} & 0\end{pmatrix},
\end{align}
with $i,j,k \in \{1,2,3\}$ and $\vec{e}_i^{\,T} = (\delta_{1i},\delta_{2i},\delta_{3i})$. They obey the following commutation relations:
\begin{align}
\begin{split}
[T_i,T_j]&=i \epsilon_{ijk} T_k,\\
[T_i^5,T_j^5]&=i \epsilon_{ijk} T_k,\\
[T_i,T_j^5]&=i \epsilon_{ijk} T_k^5,
\end{split}
\end{align}
and therefore $T_i^L=\frac{1}{2}(T_i-T_i^5)$ and $T_i^R=\frac{1}{2}(T_i+T_i^5)$ form representations of $SU(2)_L$ and $SU(2)_R$. The mesons transform under these flavor rotations as
\begin{align}
\phi \rightarrow U\phi\,,\quad V_\mu \rightarrow U V_\mu U^\dagger\,,
\end{align}
with $U=e^{i\vec{\alpha}\vec{T}+i\vec{\beta}\vec{T}^5}$ and parameters $\vec{\alpha}$ and $\vec{\beta}$. 

As already mentioned, we construct the vector-meson part of the
effective action by gauging chiral symmetry. The vector field $V_\mu$
then naturally arises as a gauge field. Its interactions with the
scalar- and pseudoscalar mesons result from minimal coupling $(D_\mu
\phi)^2/2$ with the covariant derivative $D_\mu = \partial_\mu - i g
V_\mu$. The kinetic term of the vector mesons as well as their
self-interactions arise from the field strength term $\text{tr}\,
V_{\mu\nu}V_{\mu\nu}/8$ with the field-strength tensor $V_{\mu\nu} = i
[D_\mu,D_\nu]/g$. Since the dominant contributions to the vector meson
spectral functions stem from the decay channels that involve scalar-
and pseudoscalar mesons, we have neglect the vector-meson self
interactions here. Furthermore, it has been shown in \cite{Rennecke:2015eba} that the vector mesons are always decoupled from the dynamics of the system in Euclidean space time. Our computation of the spectral functions uses the solution of the Euclidean system as input and therefore vector-meson self-interactions do not play a role for this input. However, they are important for a quantitative description of vector-meson spectral functions. As the focus of the present work is on the qualitative connection between chiral symmetry restoration and vector-meson spectral functions, we postpone a more realistic description to future work.

A soft breaking of the chiral gauge symmetry is induced by the explicit vector-meson mass term in \Eq{eq:effective_action}. However, also the Yukawa interaction terms explicitly break this symmetry. Hence, $\Gamma_k$ only possesses global chiral symmetry and, as opposed to a pure gauged linear sigma model, the $\rho$ meson does not couple to a conserved local current. An immediate consequence is that the vector-meson self energies are not purely transversal.

Lastly, we discuss the term $\Delta\Gamma_{\pi a_1}$ in \Eq{eq:effective_action}. If the mesons acquire a non-vanishing vacuum expectation value $\phi_0 = (\sigma_0,\vec{0})$ due to spontaneous chiral symmetry breaking, a contribution of the form
\begin{align}\label{eq:pia1}
-\int_x g\, \sigma_0\, \vec{a}_1^{\,\mu}\!\cdot \partial_\mu\vec{\pi} \,\subset\, \Gamma_k\,,
\end{align}
yields an off-diagonal meson propagator. This is referred to as $\pi \!-\! a_1$ mixing. We will eliminate this mixing, i.e.~diagonalize the meson propagator, by a redefinition of the $a_1$ field:
\begin{align}\label{eq:a1redef}
\vec{a}_1^{\,\mu} \longrightarrow \vec{a}_1^{\,\mu} + \frac{g \sigma_0}{m_{k,V}^2 + g \sigma_0^2}\, \partial_\mu \vec{\pi}\,.
\end{align}
If we insert this replacement into $\Gamma_k$ various new terms are generated, which we subsume in $\Delta\Gamma_{\pi a_1}$. Among the numerous new terms in the effective action only three are potentially relevant in the present work. This is rooted in the approximations we employ here. First of all, we only take the fluctuations of scalar- and pseudoscalar mesons and quarks into account. Hence, all vertices that would lead to vector mesons in the loops can be ignored. Furthermore, our construction of the effective action is based on a low-momentum expansion, i.e.~we use the lowest-order derivative expansion and hence only contributions up to second order in the derivatives have to be retained. Lastly, the scalar- and pseudoscalar meson self-interactions are defined from the effective potential $U_{k}(\phi^2)$ and are therefore momentum independent by construction. Thus, every modification of the corresponding vertices that involves space-time derivatives can also be ignored. The remaining terms yield
\begin{align}
\begin{split}\label{eq:Dpia1}
\Delta\Gamma_{\pi a_1} &= \int_x \!\bigg\{ g\, \sigma_0\, \vec{a}_1^{\,\mu}\!\cdot \partial_\mu\vec{\pi} - \frac{1}{2} \frac{g^2 \sigma_0^2}{m_{k,V}^2 + g \sigma_0^2} (\partial_\mu\vec{\pi})^2\\
&\quad -\frac{g^2 \sigma_0^2}{m_{k,V}^2 + g \sigma_0^2}\, \vec{\rho}^{\,\mu}\!\!\times\!\vec{\pi}\cdot\partial_\mu\vec{\pi} \bigg\}\,.
\end{split}
\end{align} 
The first term in the first line of this equation cancels the mixing term \Eq{eq:pia1} and thus leads to a diagonal meson propagator in the broken phase. The second term generates a non-trivial wavefunction renormalization for the pions. The effects of wavefunction renormalizations are not subject of the present analysis and are therefore postponed to future work. Hence, we drop this contribution. The term in the second line of \Eq{eq:Dpia1} modifies the $\pi\pi\rho$-vertex $\Gamma^{(3)}_{k,\pi\pi\rho}$. This has to be taken into account when we compute loop diagrams, e.g.~for the two-point function of the pion. The corresponding modified Feynman rule for this vertex can be found in \Eq{eq:vertices}. We note that the redefinition of the $a_1$ field in \Eq{eq:a1redef} involves running couplings. Thus, strictly speaking, the new $a_1$ field is explicitly renormalization group scale dependent. A self-consistent way to treat this scale-dependent field has been put forward in \cite{Rennecke:2015eba}. However, this only gives minor quantitative corrections. Since our focus is on qualitative effects, we can readily ignore this here. 

In the described model we have no mechanism included which describes the phenomenon of confinement. The Polyakov-loop within the Polyakov-quark-meson model, usually used to describe confinement in terms of thermodynamics \cite{Fukushima:2003fw,Marhauser:2008fz,Schaefer:2007pw,Skokov:2010wb}, is not able to suppress unphysical quark-antiquark decays in the confined phase, cf. \App{sec:real_imag_part}. A physically reasonable way to include confinement in such low energy models is also left to future work.

\subsection{Functional renormalization group and flow equations}\label{subsec:frg}
\begin{figure}[h]
	\includegraphics[width=0.33\textwidth]{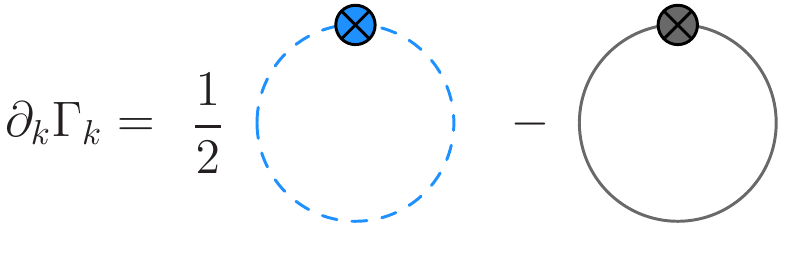}
	\caption{(color online) Flow equation for the effective action of the model defined by \Eq{eq:effective_action} in diagrammatic form. The dashed blue line is associated with propagators of the $\sigma$- and $\pi$ meson, the solid black line with the quark propagator, while the crossed circles represent the regulator insertions $\partial_k R_k$.}\label{fig:flow_eq_qm-model}
\end{figure}

The FRG provides a powerful, non-perturbative tool to investigate the transition from microscopic to macroscopic scales and is widely used in statistical physics and quantum field theory. The change of the effective average action $\Gamma_k$ with momentum scale $k$ is described by the Wetterich equation \cite{Wetterich:1992yh}
\begin{align}
\label{eq:wetterich}
\partial_k \Gamma_k = \frac{1}{2} \Tr \left[\partial_k R_k^{\phi}~ \left(\Gamma^{(2)}_k[\phi]+R_k^{\phi}\right)^{-1}\right].
\end{align}
Here we apply this concept to the low-energy model introduced in
\Sec{subsec:model}, see \Fig{fig:flow_eq_qm-model} for a diagrammatic
representation of the resulting flow equation. The Wetterich equation
then turns into a flow equation for the effective mesonic potential
$U_k$, see also \Eq{eq:flow_eq_eff_pot} for its explicit form. We note that for isospin symmetric matter with an equal number of up and down quarks, the isovector rho and a1 mesons do not contribute to the effective potential.

Flow equations for $n$-point functions are in general obtained by taking $n$ functional derivatives of the Wetterich equation with respect to certain fields. The flow equations then naturally contain up to $n+2$-point functions which leads to an infinite set of coupled equations. In order to solve this system, one has to introduce truncations. In the present work we extract $3$- and $4$-point functions appearing on the RHS of the flow equations from the ansatz of the effective average action, \Eq{eq:effective_action}, as done for example in \cite{Tripolt2014,Tripolt2014a}. An improved truncation taking into account full momentum dependent vertices and solving the flow equations for $2$-point functions iteratively is left to future studies.

\begin{figure*}[t]
	\includegraphics[width=0.9\textwidth]{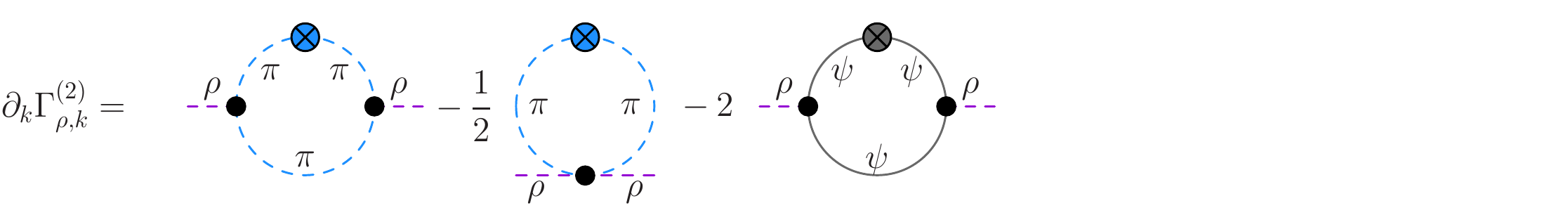}\\
	\includegraphics[width=0.9\textwidth]{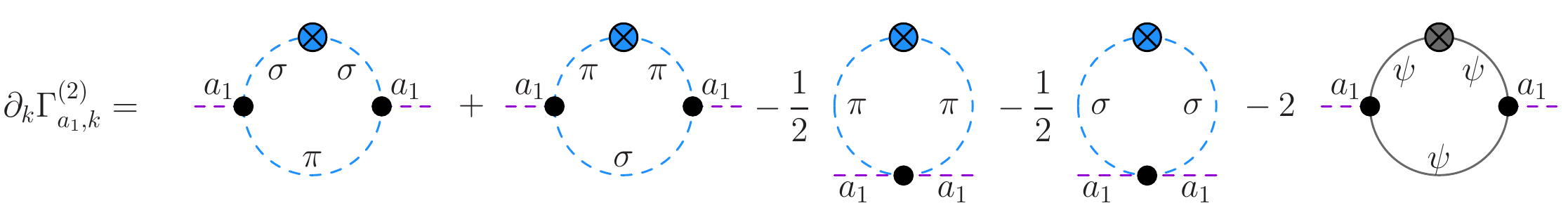}
	\caption{(color online) Flow equations for the $\rho$ and $a_1$ two-point functions in diagrammatic form. Vertices are indicated by black filled dots, regulator insertions by crossed circles. The color of the lines and regulators represents the type of field: blue for scalar and pseudoscalar mesons, black for fermions and purple for vector mesons.}\label{fig:flow_equations_two-point}
\end{figure*}

Following this strategy, we obtain flow equations for the $\rho$ and $a_1$ two-point functions as a closed system of equations
\begin{alignat}{2}
\label{eq:flow_eq_two-point_functions}
&\partial_k \Gamma_{k,\rho}^{(2)}(p) &&= J^{\pi\pi}_{k,\rho}(p)-\frac{1}{2}I^{\pi}_{k,\rho}-2 J^{\psi\bar{\psi}}_{k,\rho}(p),\\
&\partial_k \Gamma_{k,a_1}^{(2)}(p) &&= J^{\sigma\pi}_{k,a_1}(p)+J^{\pi\sigma}_{k,a_1}(p)\nonumber\\
& &&\quad-\frac{1}{2}I^{\pi}_{k,a_1}-\frac{1}{2}I^{\sigma}_{k,a_1}-2 J^{\psi\bar{\psi}}_{k,a_1}(p),
\end{alignat}
where momentum-dependent and momentum-independent loop functions appear, see \App{sec:flow_equations} and \App{sec:loop_functions} for the projected flow equations and explicit expressions, including the vertices.
The diagrammatic form of these equations is illustrated in \Fig{fig:flow_equations_two-point}.

The flow equations are then solved by starting with the microscopic theory at a UV-cutoff and then integrating out momentum shells down to the scale $k$. The effective average action $\Gamma_k$ as well as the mesonic two-point functions $\Gamma^{(2)}_k$ then incorporate fluctuations with momenta larger than the RG-scale $k$. For technical details we refer to \Sec{subsec:numerics} and \App{sec:flow_equations}.

\section{Results and Discussion}\label{sec:results}

\subsection{Numerical implementation}\label{subsec:numerics}
In order to solve the flow equations for the retarded two-point
functions of the $\rho$ and $a_1$ meson, we first solve the flow
equation for the effective potential $U_k$,
cf.~App.~\ref{sec:flow_equations}. This is done by discretizing the
effective potential in $\sigma$-field direction which yields a set of
coupled ordinary differential equations which can then be solved with
common methods. The parameters for the effective potential at the UV
scale $\Lambda = 1500$ MeV,
\begin{equation}
U_{\Lambda} = b_1 \phi^2+b_2
\phi^4\, ,
\end{equation}
the explicit symmetry-breaking constant $c$ as well as
the scalar and vector Yukawa couplings are given in
Tab.~\ref{tab:parameters}. They are chosen such that
in the vacuum we have, with a constituent quark mass of \mbox{$m_{\psi} =
  300$ MeV} at the IR scale \mbox{$k_{\text{IR}} =
  40$ MeV}, phenomenologically reasonable values for the
masses of pion and $\sigma$ meson (corresponding to the $f_0(500)$
resonance) and the pion decay constant (here identified with the global
minimum of the $\sigma$-field at $\sigma_0$): \mbox{$m_{\pi} =  140$ MeV},
\mbox{$m_{\sigma} =  557$ MeV}, 
\mbox{$f_{\pi} \equiv \sigma_0 =  93.0$ MeV}. 

Note that the meson masses here are Euclidean curvature masses,
i.e.~the mass parameters that determine the curvatures of the mesonic
effective potential which agree with the zero-momentum limits of
the corresponding Euclidean mesonic 2-point functions. They can differ
from the physical masses which are the so-called pole masses, i.e.~the
zeros of the analytically continued 2-point functions at time-like
momenta. Apart from potential wavefunction renormalization factors,
beyond the leading order derivative-expansion employed here, the
differences will be small for the pion, for example, because its
zero-momentum correlator is of course dominated from the well-isolated
pion pole. For heavier and not so well-isolated bound states and resonances
such as the $\rho$ and $a_1$ mesons these differences will be important,
however.             

\begin{table}[b]
	\centering
	\begin{tabular}{C{1.4cm}|C{0.6cm}|C{1.5cm}|C{1.4cm}|C{0.6cm}|C{1.7cm}}
		$b_1 \text{~[MeV}^2\text{]}$ & $b_2$ & $c\text{~[MeV}^3\text{]}$ & $h_S=h_V$&$g$ & $m_{\Lambda,V}\text{~[MeV}\text{]}$\\
		\hline\hline
		857300  & 0.2 & 1.8228$\cdot 10^6$ &3.226 & 11.4 & 1450
	\end{tabular}
	\caption{Parameter set used in this work.}
	\label{tab:parameters} 
\end{table}

The Euclidean curvature mass of the $\rho$ meson is given by the
vector-meson mass parameter, i.e.~their quadratic coupling $m^2_{k,V}$
in the effective action in Eq.~(\ref{eq:effective_action}),
  cf.~App.~\ref{sec:flow_equations}.   
By using the RG flow of the effective potential and its derivatives as
input, the flow equation for $m^2_{k,V}$,  Eq.~(\ref{eq:flow_mV}), can
be solved. The initial condition for  $m^2_{\Lambda,V}$ at the UV scale
 $\Lambda = 1500$ MeV, as well as the scale-independent gauge coupling
$g$ as chosen in Tab.~\ref{tab:parameters} then result in the pole
masses $m_{\rho}^p$ and $m_{a_1}^p$, for the resonances here defined as the 
zero-crossing of the real part of the retarded two-point function,
cf.~App.~\ref{sec:real_imag_part}, in the vacuum: \mbox{$m_{\rho}^p =
  789.3$ MeV} and \mbox{$m_{a_1}^p =  1274.7$ MeV}. They thus
reproduce the physical masses of $\rho$ and $a_1$ reasonably well.
Since fine-tuning the UV parameters to adjust these pole masses to the
physical ones more precisely is a rather tedious task, we are content
with this level of agreement for our qualitative study here.
For comparison, the corresponding Euclidean curvature masses with the same UV
parameters result as \mbox{$m_{\rho}=1298.3$ MeV} and \mbox{$m_{a_1}=1676.3$
  MeV}. Parts of the discrepancy between curvature and
pole masses should be compensated by the inclusion of wavefunction
renormalization factors, i.e.~by going to higher orders in the
gradient expansion. We reiterate 
however, that there is no a priori reason for the two to agree.
In particular, the Euclidean curvature masses do not have a direct
  physical meaning and should rather be seen as parameters that determine the
physical pole masses. 

In a last step the flow equations for the real and the imaginary part of the retarded two-point functions are solved at the grid point of the IR minimum $\sigma_0$, from which the spectral functions can be obtained as described in App.~\ref{sec:analytic_continuation}.

\subsection{Phase structure and curvature masses}\label{subsec:masses}

\begin{figure}[t]
	\includegraphics[width=0.48\textwidth]{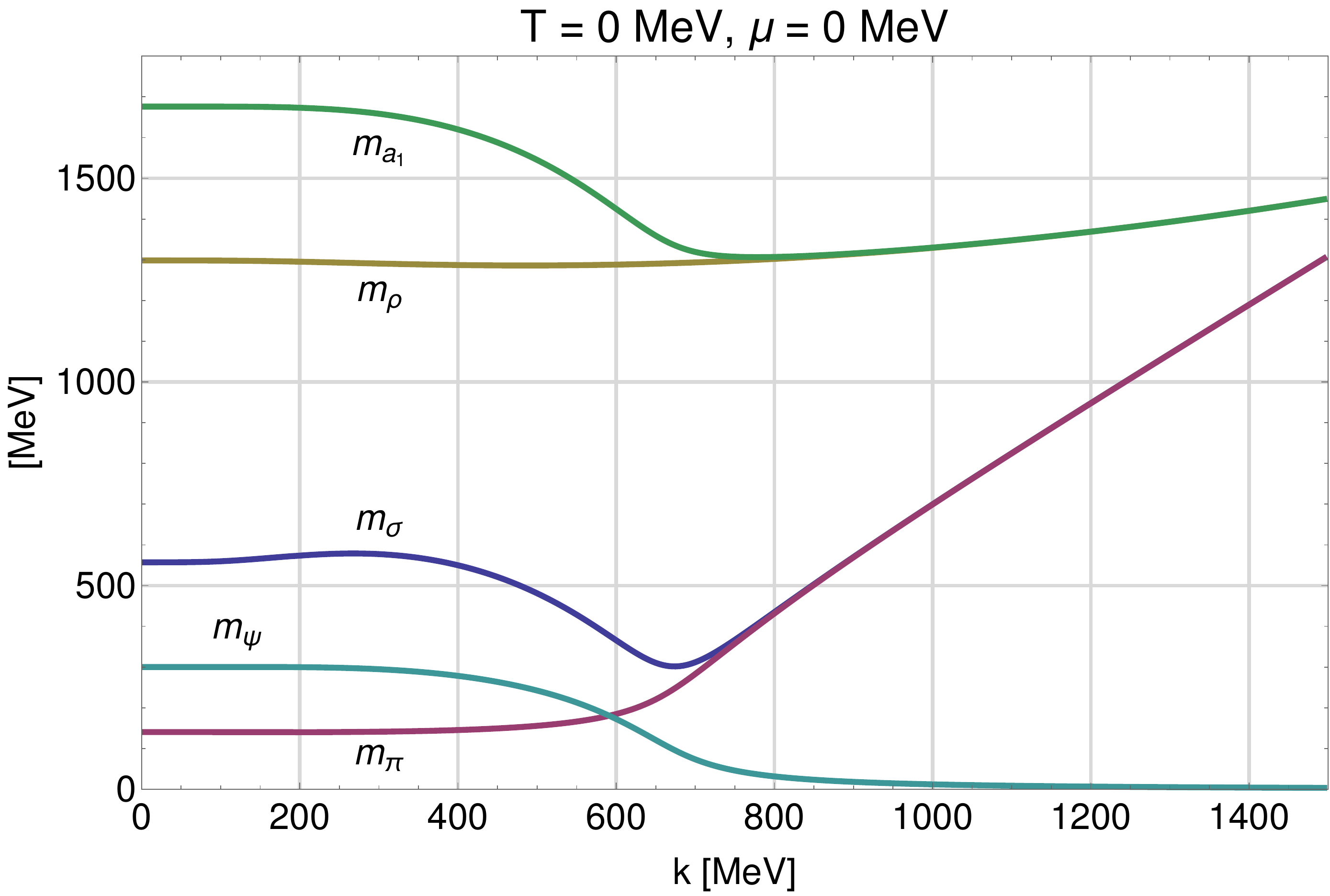}
	\caption{(color online) Flow of the scale-dependent Euclidean curvature masses of mesons and the constituent quark mass with the RG scale $k$ at \mbox{$T=0$ MeV} and \mbox{$\mu = 0$ MeV}.}\label{fig:scale_masses} 
\end{figure}

\begin{figure}[b!]
	\includegraphics[width=0.49\textwidth]{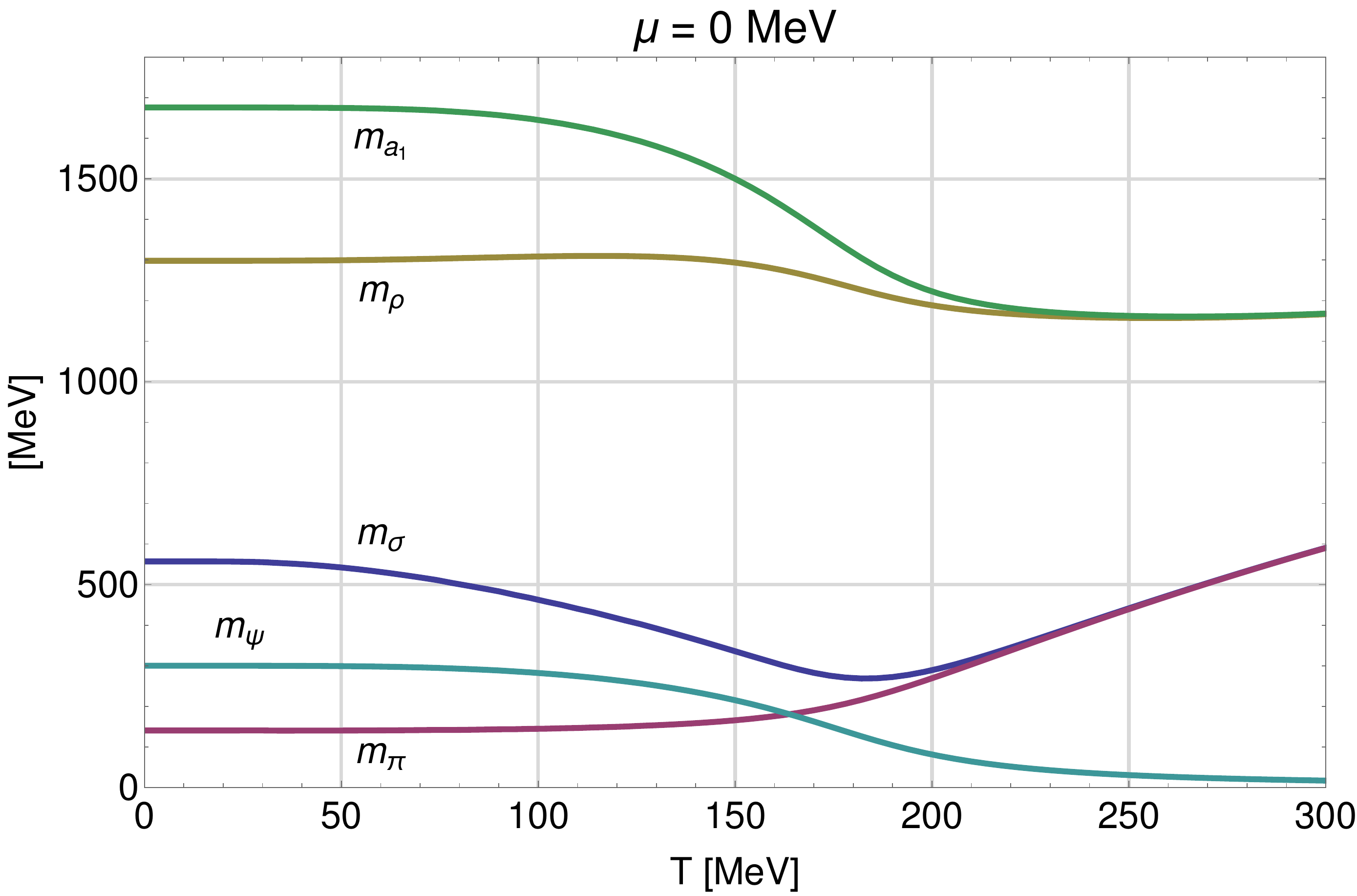}
	\caption{(color online) Euclidean curvature masses of  mesons
		and constituent quark mass at the IR scale vs.~temperature
		at $\mu=0$~MeV.}\label{fig:euclidean_masses_T} 
\end{figure}

In this section we briefly discuss the flow of the Euclidean curvature
masses as well as the phase structure and the $T$- and
$\mu$-dependence of the quark and meson masses which serve as an
important input for the computation of the spectral functions. 

\begin{figure}[t]
	\includegraphics[width=0.99\columnwidth]{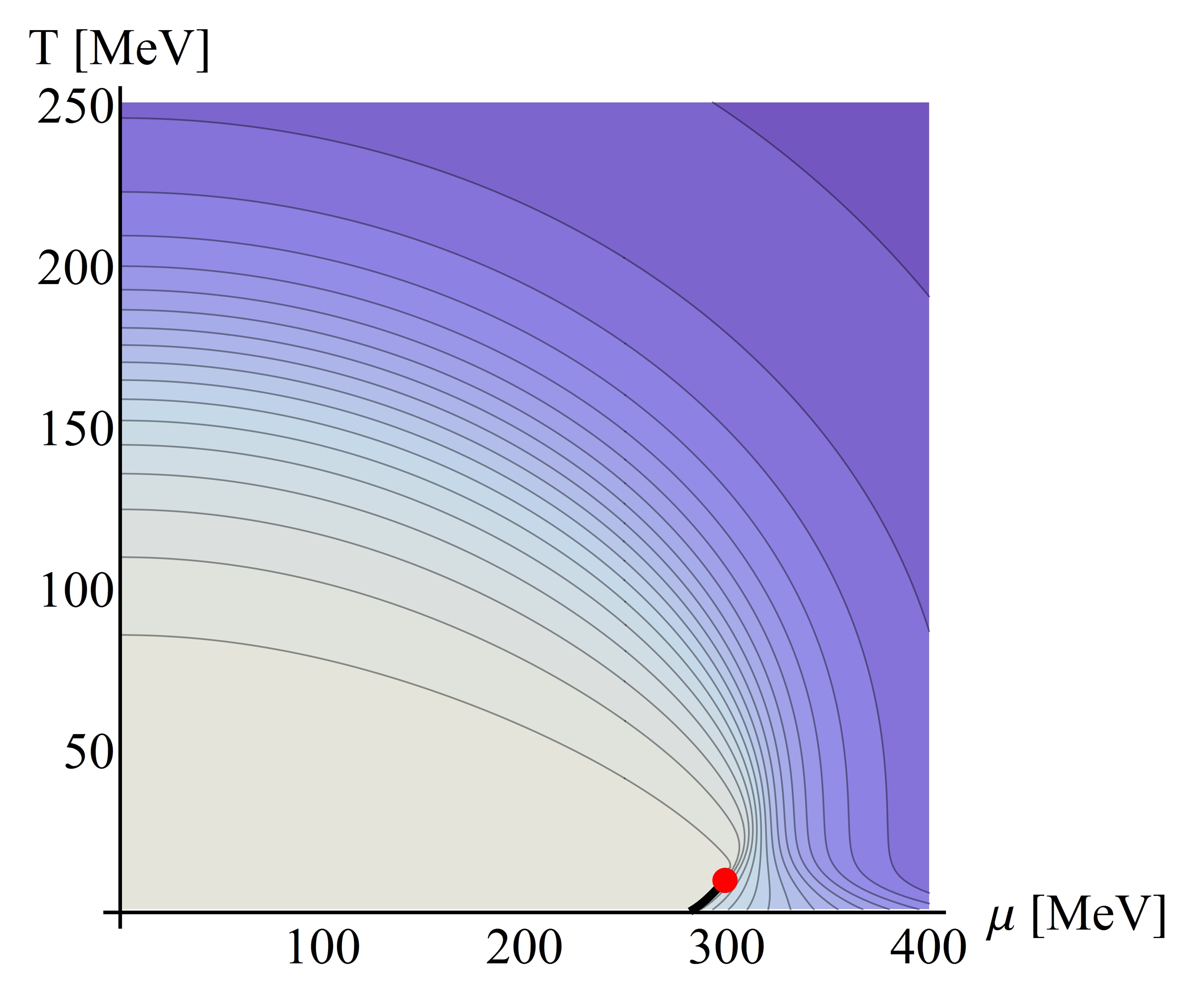}\llap{\makebox[4cm][l]{\raisebox{3.3cm}{\includegraphics[height=3.5cm]{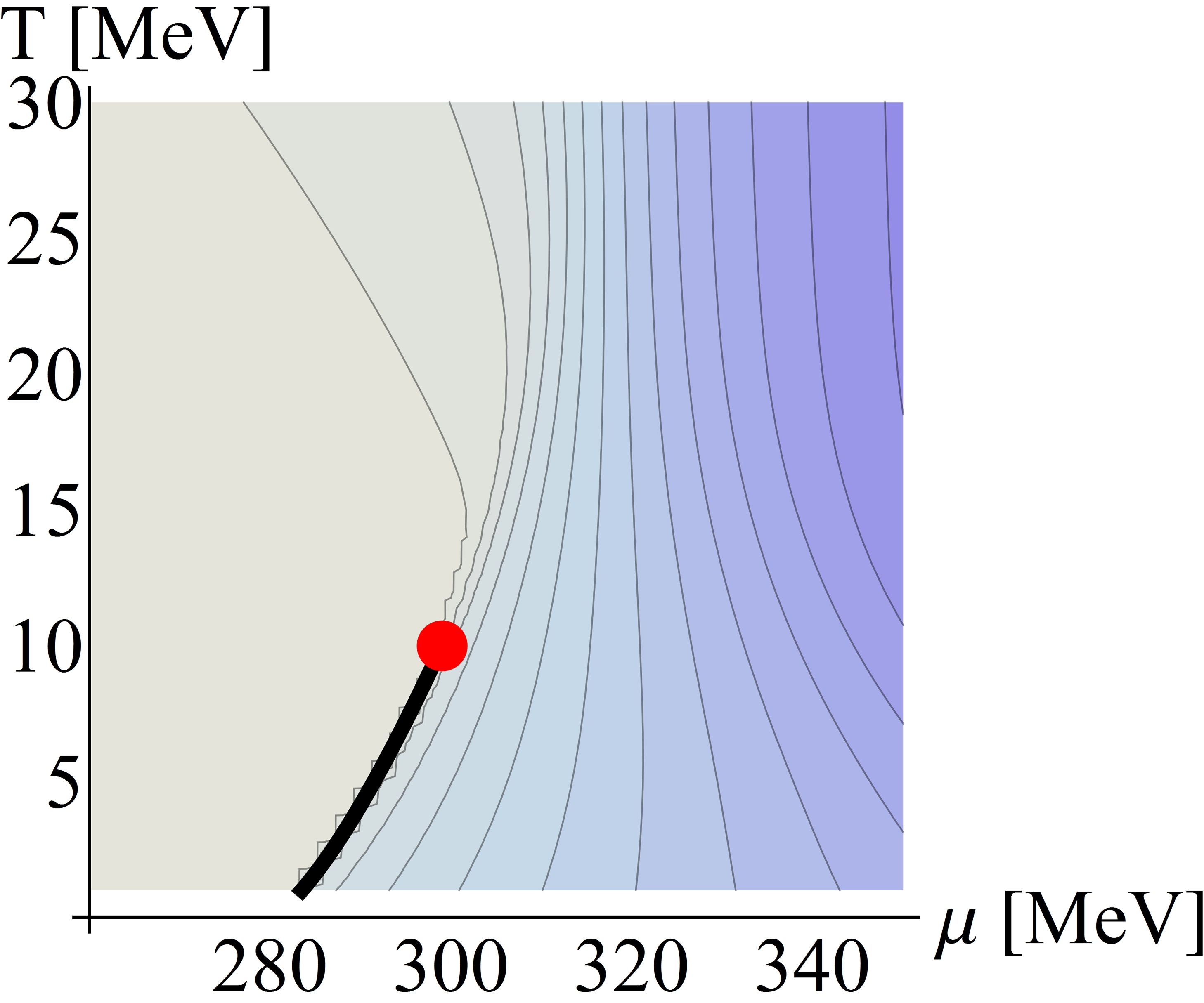}}}}
	\caption{(color online) Phase diagram of the quark-meson model as a contour plot of the order parameter for chiral symmetry $\sigma_0(T,\mu)$. The value of $\sigma_0(T,\mu)$ decreases with increasing temperature and chemical potential as indicated with a darker color. The CEP is indicated as a red dot, whereas the first-order phase boundary is indicated by a black line.}
	\label{fig:phase_diagram} 
\end{figure}

\begin{figure}[b]
	\includegraphics[width=0.49\textwidth]{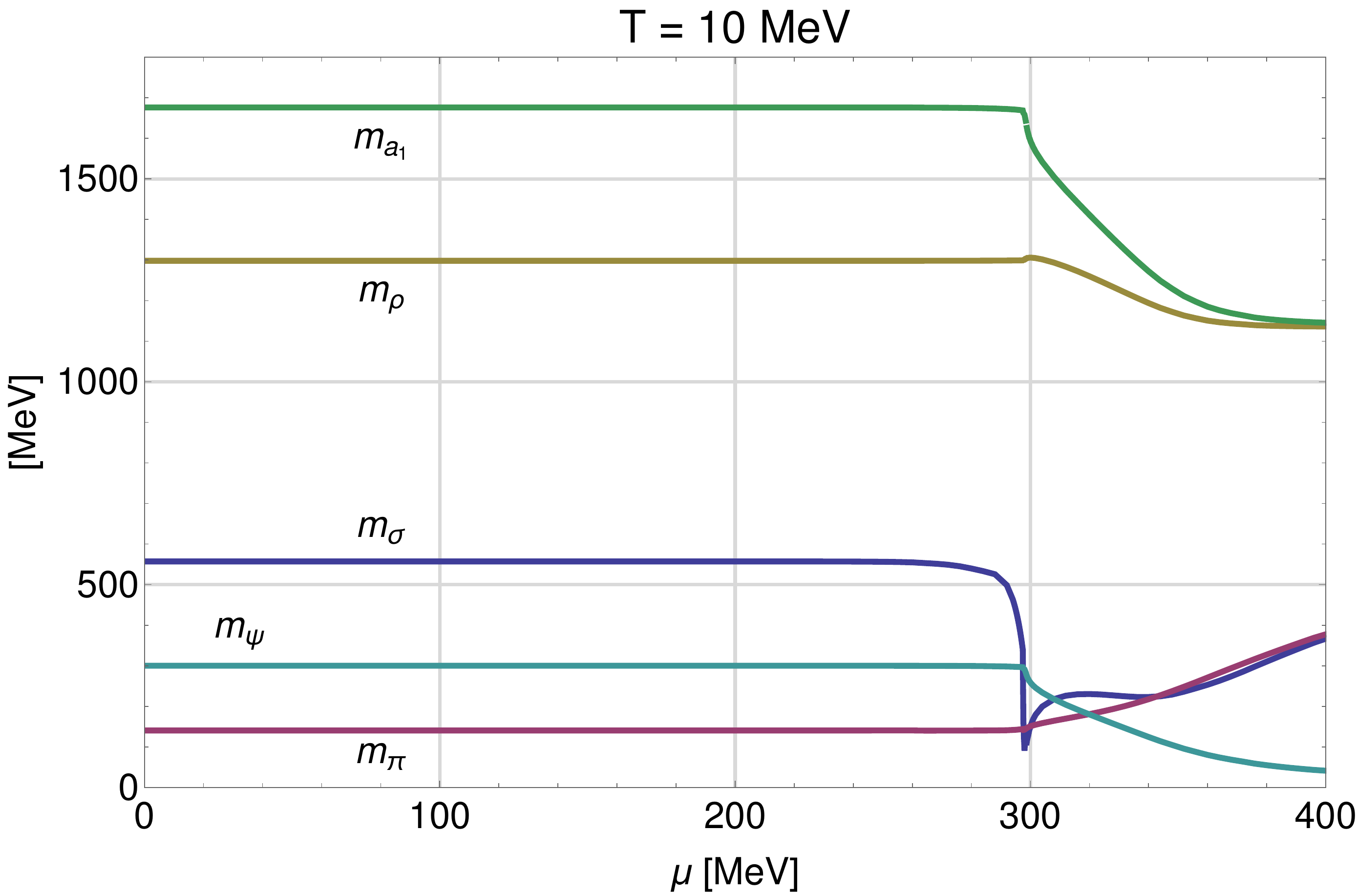}
	\caption{(color online) Euclidean curvature masses of mesons
		and constituent quark mass vs.~chemical potential at
		\mbox{$T=10$ MeV}.}\label{fig:euclidean_masses_mu} 
\end{figure}

The RG flow of the scale-dependent Euclidean curvature masses defined
in App.~\ref{sec:flow_equations} is shown in
Fig.~\ref{fig:scale_masses}. Starting in the chirally restored phase
at the UV cutoff \mbox{$\Lambda=1500$ MeV}, the masses of the chiral
partners are degenerate. When lowering the RG scale, both the $\rho$
and the $a_1$ meson masses slightly drop, in agreement with the
QCD-based study in \cite{Rennecke:2015eba}. Entering the regime
where chiral symmetry gets dynamically broken at around
\mbox{$k_{\chi_{\text{SB}}} \approx 700$ MeV}, the mass of the $a_1$
meson begins to rise, whereas the $\rho$ meson mass remains
approximately constant during the complete flow. We notice that the
vector-meson masses are always larger than the scale $k$ and are hence
always decoupled from the flow, which has also been observed in
\cite{Rennecke:2015eba}. The dynamics in Euclidean space-time are
therefore completely determined by the pion and the $\sigma$ meson as
well as the quarks. The overall behavior with the RG scale $k$
qualitatively resembles the temperature dependence of quark and meson
masses at \mbox{$\mu = 0$ MeV} in Fig.~\ref{fig:euclidean_masses_T}.

The phase diagram of the model we use, which is a quark-meson model on
the level of the effective potential, is depicted in
Fig.~\ref{fig:phase_diagram}, see also
\cite{Tripolt2014,Schaefer:2004en} for earlier studies on the
quark-meson model. It is obtained by the location of the global
minimum of the effective potential at the IR scale
$\sigma_0\equiv\sigma_0(T,\mu)$. With the parameters given in
Tab.~\ref{tab:parameters} we find a critical endpoint at around
\mbox{$(\mu_{\text{CEP}},T_{\text{CEP}})\approx (298,10)$ MeV}, which
divides a crossover region from a first-order phase transition at
lower temperatures. We note that the slope, $dT/d\mu$, of the
first-order line is very different than the one observed in mean-field
studies, see e.g.~\cite{CarignanoBuballaSchaefer2014}. In fact, the
regime to the right of the first-order line, i.e.~at large chemical potentials and low temperatures, is likely to be dominated by an inhomogeneous ground state which leads to unphysical effects like a negative entropy density in the present truncation. We therefore avoid this regime in the following and refer to \cite{TripoltSchaeferSmekalEtAl2016} for further details.

The same Euclidean curvature masses of the mesons, plotted together with the
constituent quark mass over temperature at \mbox{$\mu = 0$ MeV} in
Fig.~\ref{fig:euclidean_masses_T}, are shown along the $\mu$-axis
at a constant temperature of \mbox{$T=10$ MeV}  across the CEP 
 in Fig.~\ref{fig:euclidean_masses_mu}. They behave as expected in a
 model based on chiral symmetry. For vanishing chemical potential, the
 Euclidean curvature masses of the chiral partners $m_{\sigma}$,
 $m_{\pi}$ and $m_{\rho}$, $m_{a_1}$ become degenerate at high
 temperatures, \mbox{$T\gtrsim 200$ MeV}. The quark mass $m_{\psi}$
 decreases, indicating the gradual restoration of chiral symmetry. For
 a fixed temperature of \mbox{$T=10$ MeV} the masses do not really
 change over a wide range of chemical potential, as expected from the
 Silver Blaze property \cite{Cohen2003}. Near the CEP at around
 \mbox{$\mu_{\text{CEP}} \approx 298$ MeV}, the sigma mass drops
 significantly as expected at this second-order phase transition. In
 addition, the chiral condensate as well as the vector-meson masses
 decrease when crossing the CEP. For very high chemical potentials the
 masses of the chiral partners coincide again and the quark mass
 decreases, similar to the case of high temperature and vanishing
 chemical potential. 

\subsection{In-medium spectral functions at $|\vec{p}|=0$}\label{subsec:spectral_function}

\begin{figure}[t]
	\includegraphics[width=0.49\textwidth]{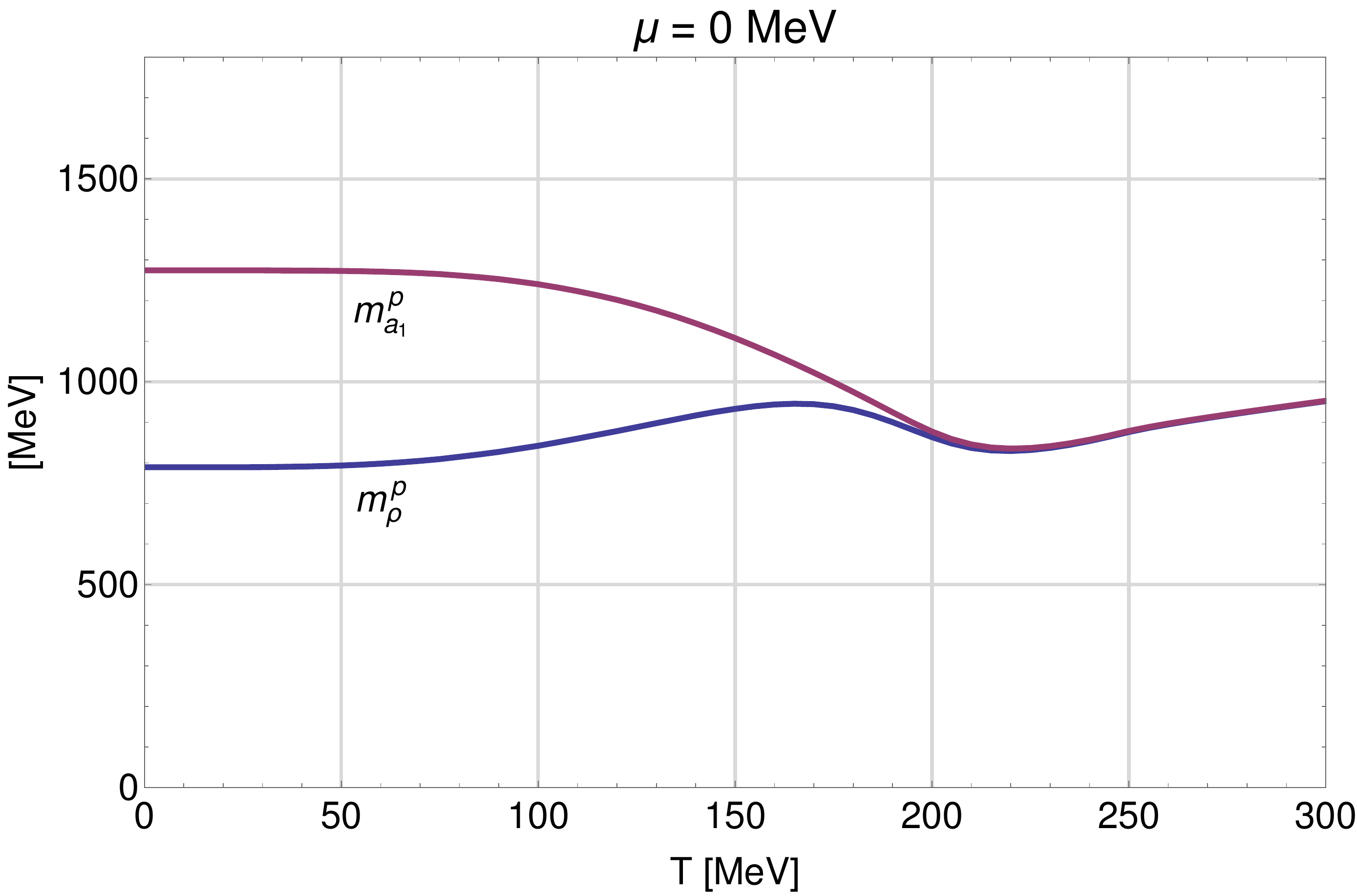}
	\caption{(color online) Pole masses of $\rho$ and $a_1$ meson vs.~temperature at $\mu=0$ MeV.}\label{fig:pole_masses}
\end{figure}

Before turning to the $\rho$ and $a_1$ spectral functions at finite temperature and chemical potential, in this subsection for vanishing external spatial momentum, $\left|\vec{p}\right| = 0$, we will discuss the temperature dependence of the physically relevant vector-meson pole masses. They are obtained from the zero-crossing of the real part of the 2-point functions and are shown in Fig.~\ref{fig:pole_masses} vs.~$T$ at $\mu=0$. 

At $T=0$ chiral symmetry is broken and the pole masses assume the vacuum values of \mbox{$m_{\rho}^p = 789.3$ MeV} and \mbox{$m_{a_1}^p = 1274.7$ MeV} for the UV-parameters of Tab.~\ref{tab:parameters}. With increasing temperature the difference between the pole masses decreases until they become degenerate at $T\approx 200$~MeV, i.e.~at about the same temperature as the Euclidean curvature masses. The observed behavior supports the ``melting-$\rho$-scenario'', where the $\rho$  meson mass remains almost stable and the $a_1$ mass shifts towards the mass of the $\rho$ meson \cite{Rapp:2009yu,Hohler:2015iba}.

\begin{figure*}
\begin{center}
	\includegraphics[width=0.45\textwidth]{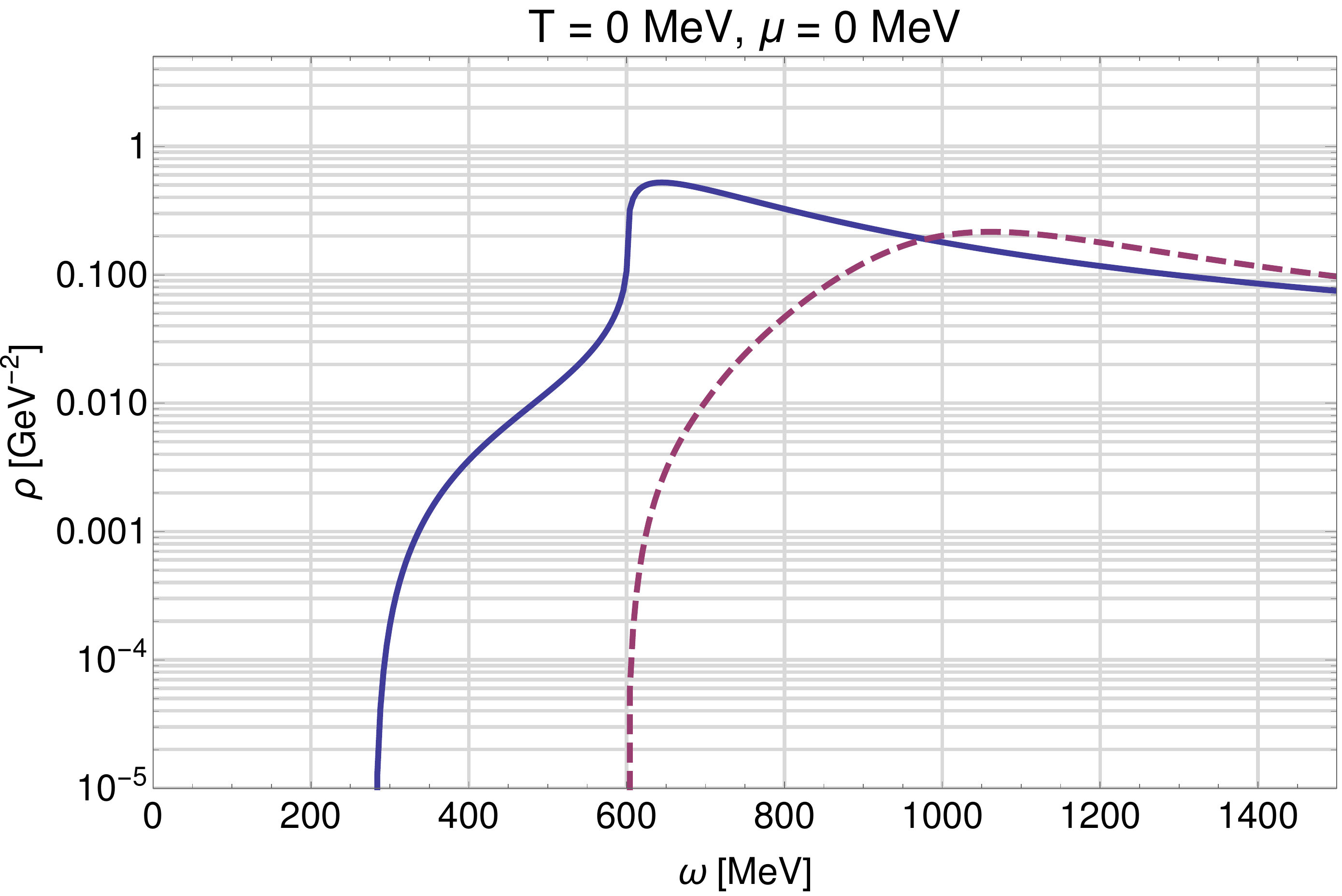}\hspace{.5cm}
	\includegraphics[width=0.45\textwidth]{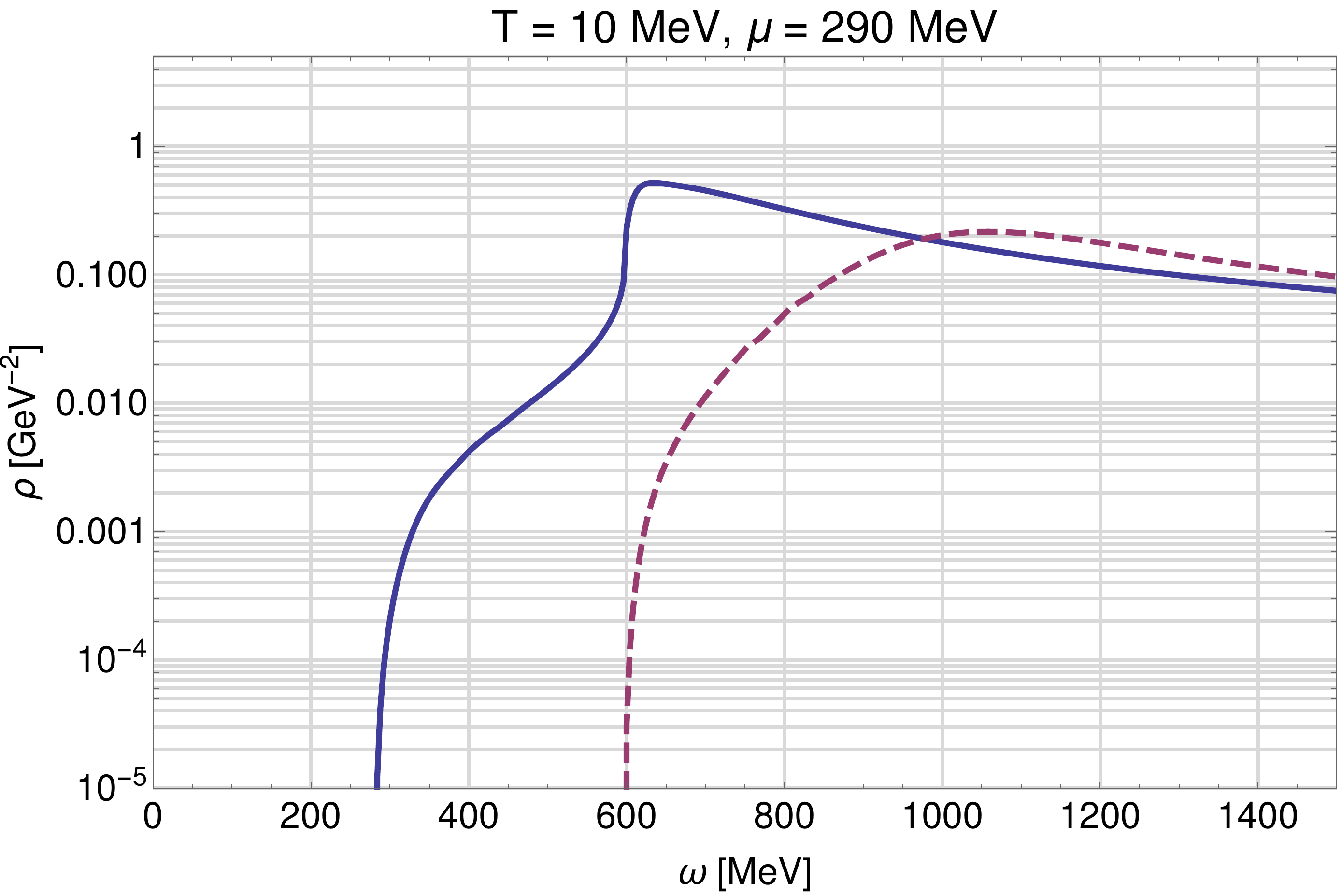}\\\vspace{5mm}
	\includegraphics[width=0.45\textwidth]{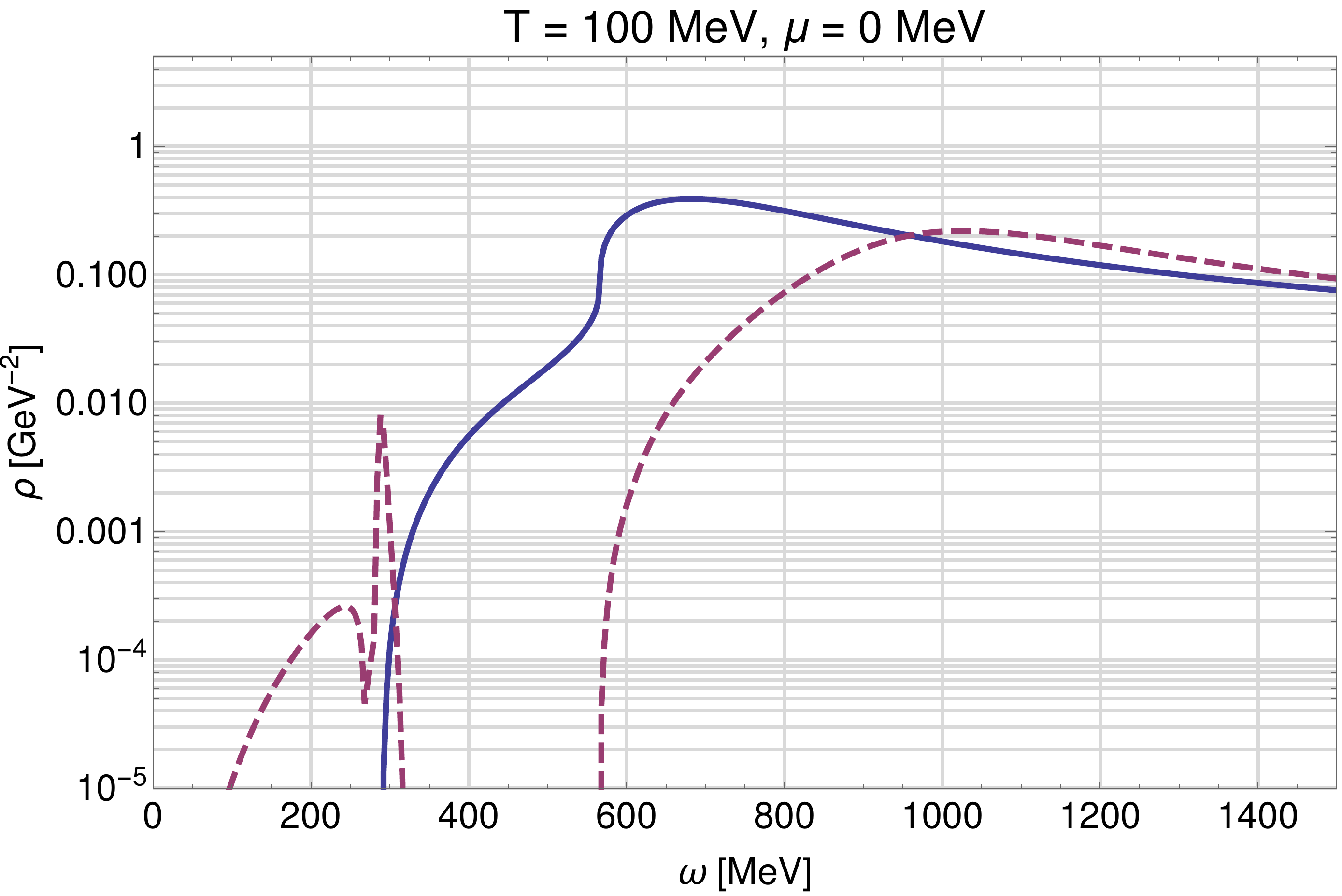}\hspace{.5cm}
	\includegraphics[width=0.45\textwidth]{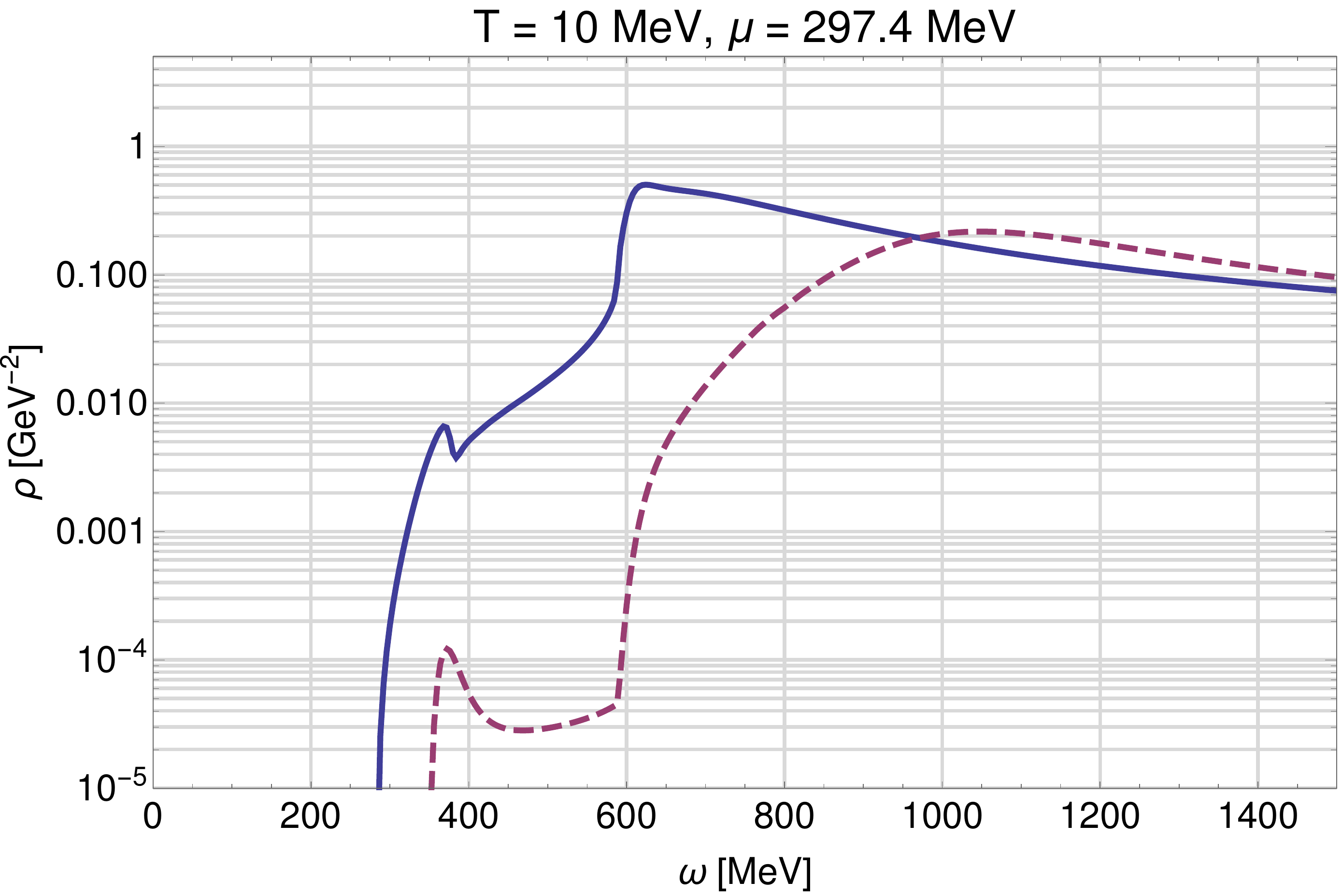}\\\vspace{5mm}
	\includegraphics[width=0.45\textwidth]{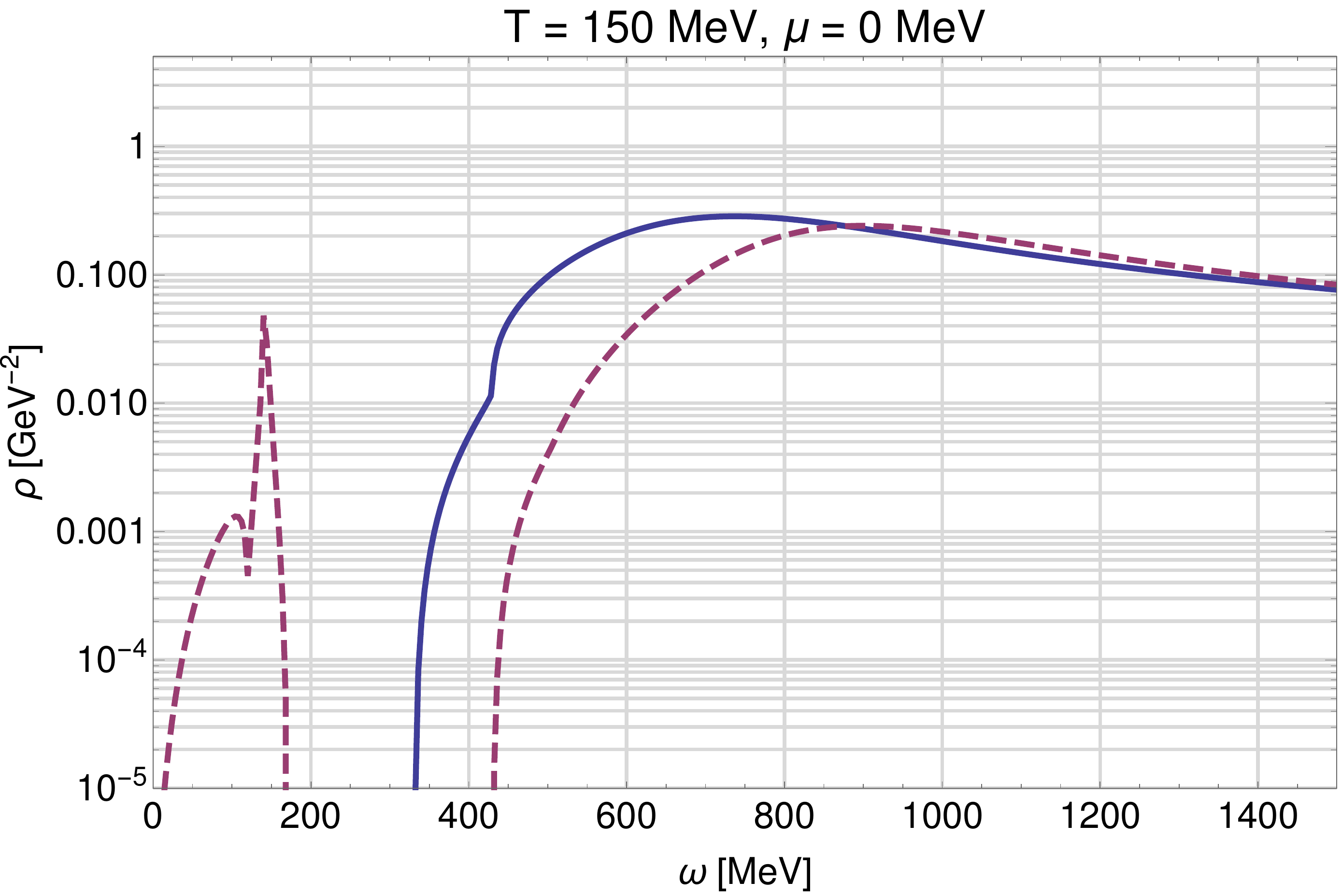}\hspace{.5cm}
	\includegraphics[width=0.45\textwidth]{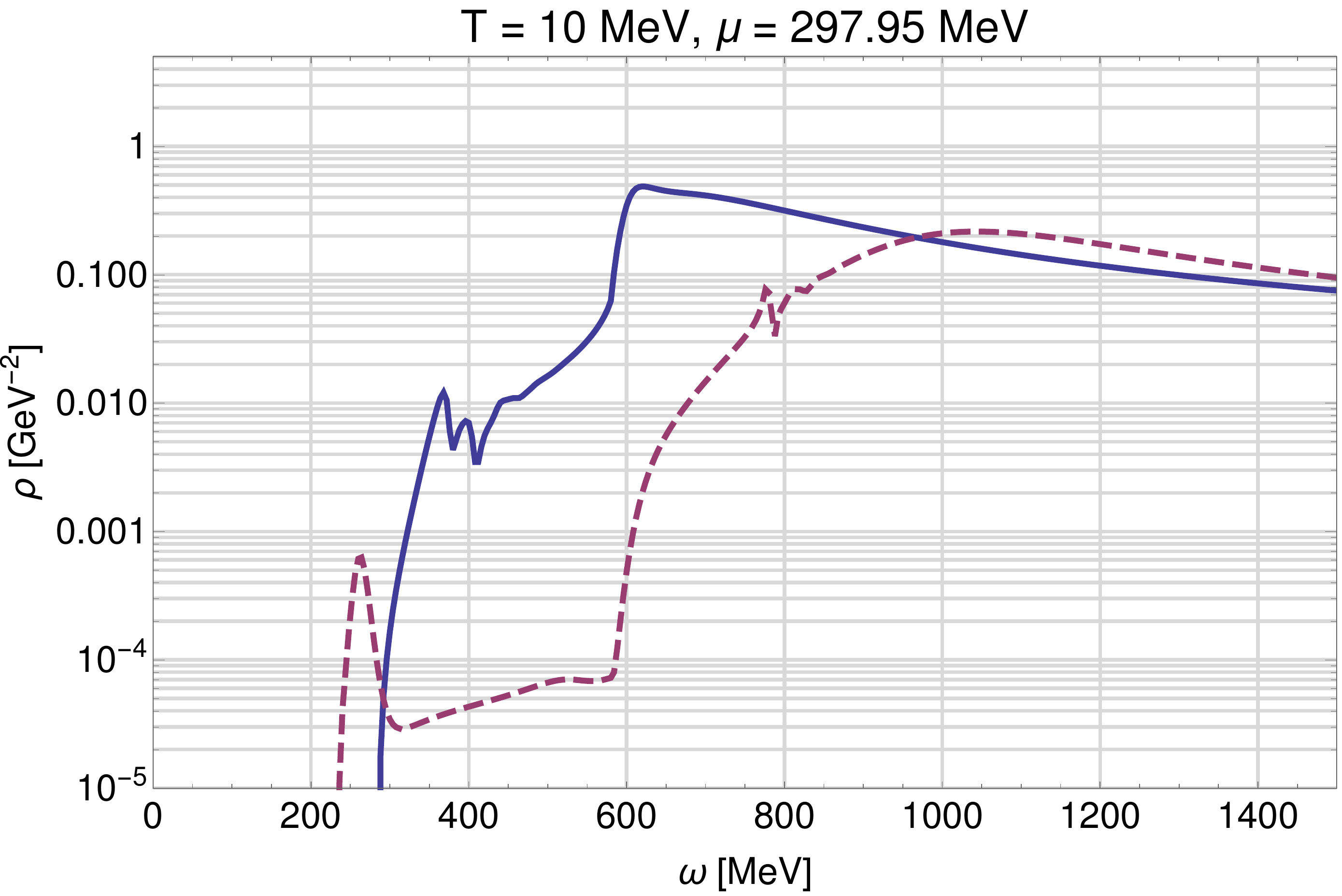}\\\vspace{5mm}
	\includegraphics[width=0.45\textwidth]{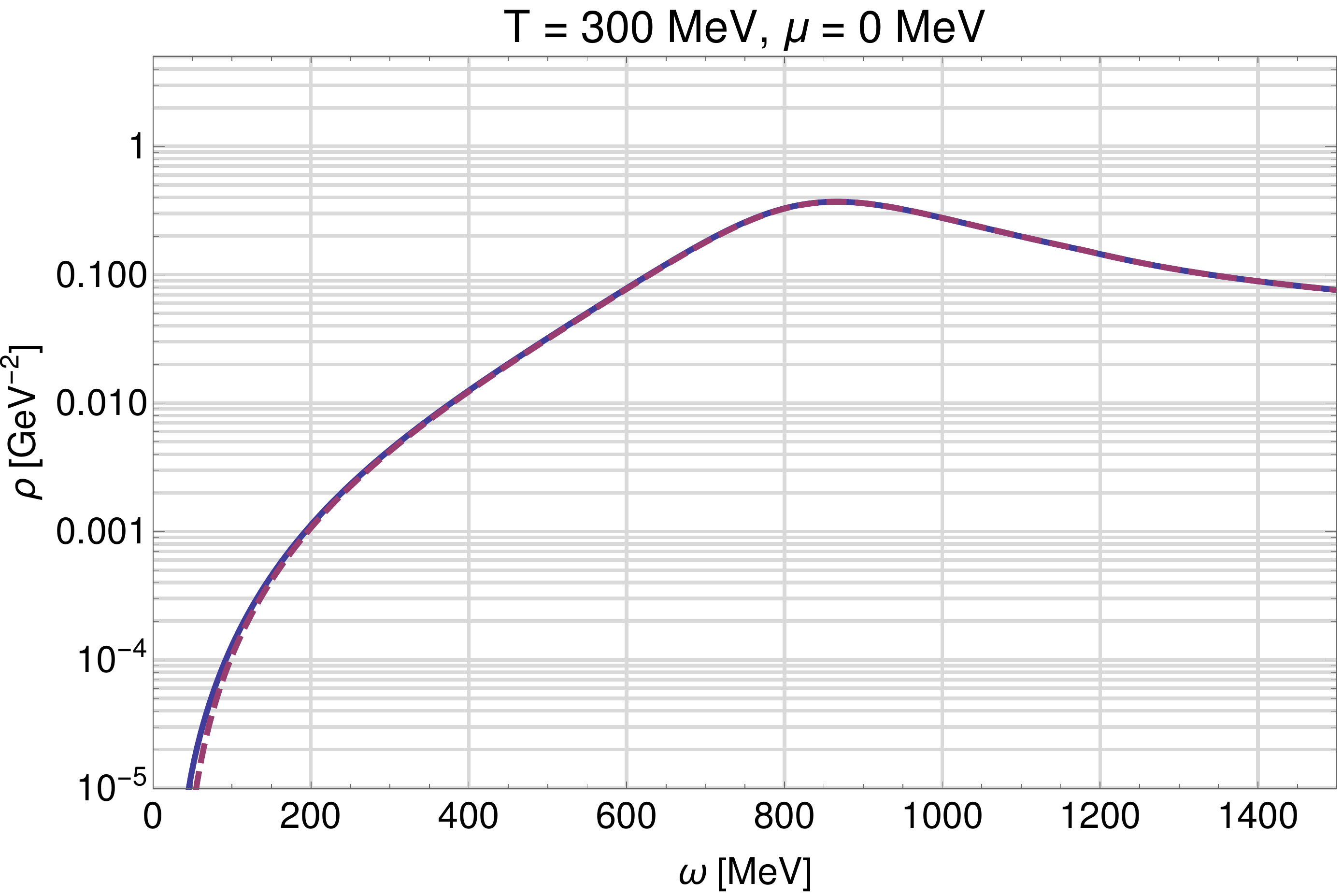}\hspace{.5cm}
	\includegraphics[width=0.45\textwidth]{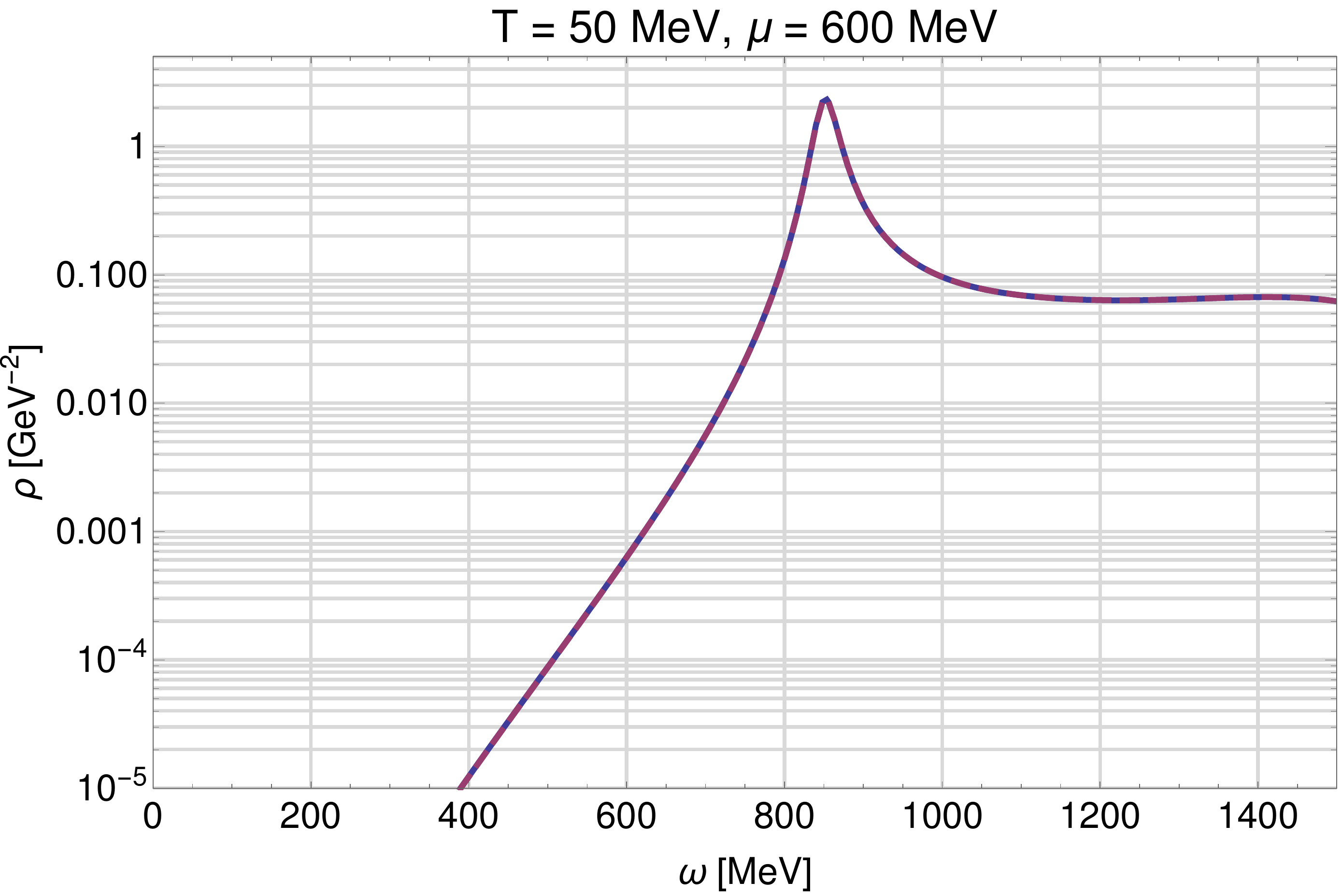}
	\caption{(color online) Spectral functions of $\rho$ (solid blue) and $a_1$ (dashed purple) meson vs.~external energy $\omega$ at $\left|\vec{p}\right| = 0$ shown for increasing temperature at \mbox{$\mu=0$ MeV} (left column) and for increasing chemical potential at \mbox{$T=10$ MeV}, towards the CEP (right column). For the last plot on the right-hand side we chose $T=50$~MeV and $\mu=600$~MeV in order to avoid a thermodynamically problematic regime in the phase diagram, see Sec.~\ref{subsec:masses} for details.}\label{fig:spectral_functions_temp_mu}
\end{center}
\end{figure*}

In order to exhibit the $T$- and $\mu$-induced modifications more clearly, Fig.~\ref{fig:spectral_functions_temp_mu} shows logarithmic plots of the $\rho$ and $a_1$ spectral functions for vanishing chemical potential (left column) and for a fixed temperature of \mbox{$T=10$ MeV} along the $\mu$-axis towards the CEP (right column). At $T=0$~MeV, the \mbox{$\rho^*\rightarrow \pi+\pi$} threshold gives rise to a non-vanishing value of the $\rho$ spectral function for \mbox{$\omega \gtrsim 280$ MeV}. For \mbox{$\omega \gtrsim 600$ MeV} the decay into quark-antiquark pairs becomes energetically possible and gives rise to another threshold in the spectral function. The spectral function of the $a_1$ meson exhibits the first threshold at \mbox{$\omega \approx 600$ MeV}, when the quark-antiquark decay becomes possible and the width then increases due to the mesonic decay channel \mbox{$a_1^*\rightarrow \pi+\sigma$}. In both spectral functions, the quark-antiquark decay strongly suppresses the pole mass peaks, see also App.~\ref{sec:real_imag_part}.

\begin{figure*}[t]
	\includegraphics[width=\columnwidth]{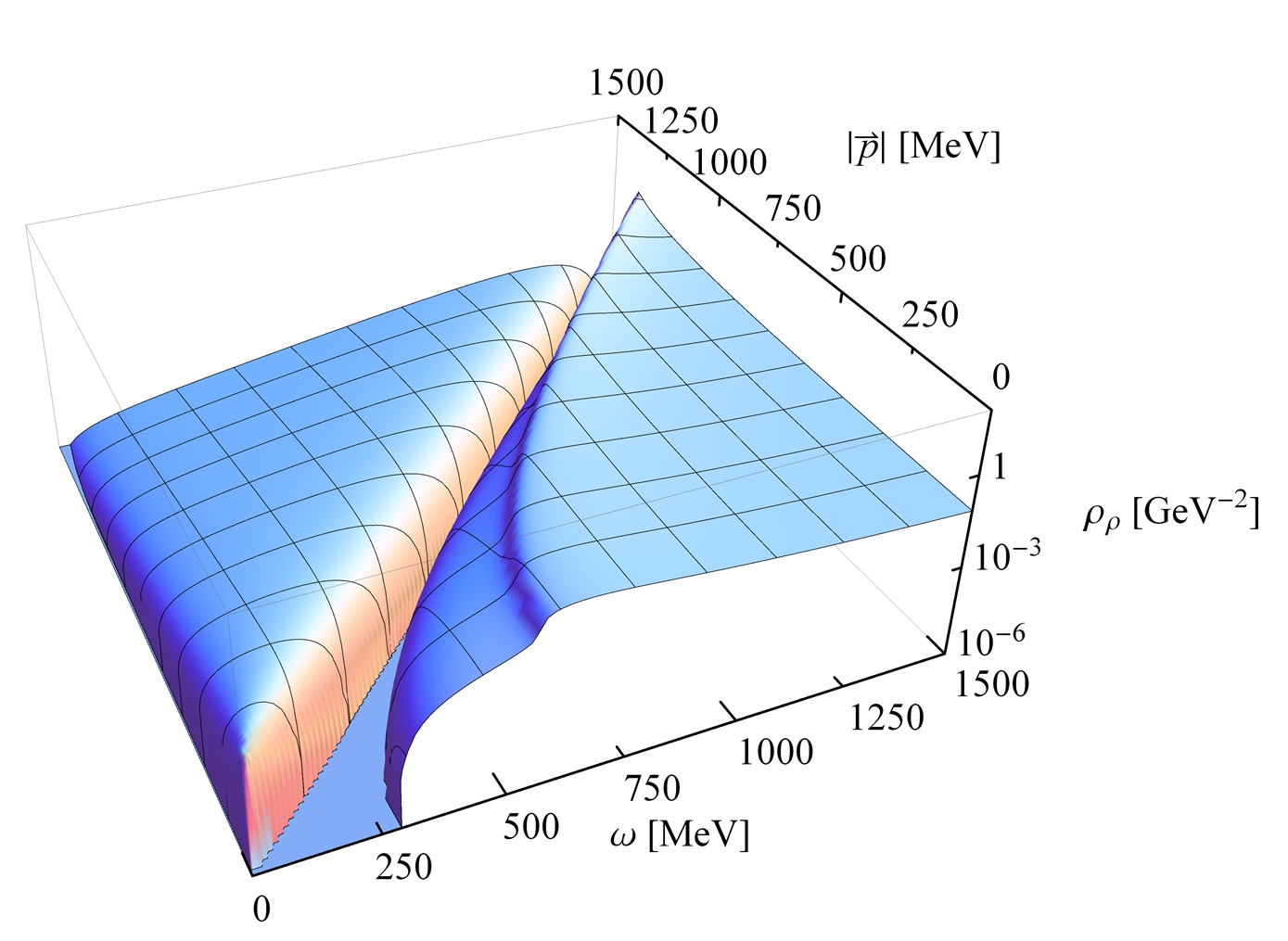}\hspace{3mm}
	\includegraphics[width=\columnwidth]{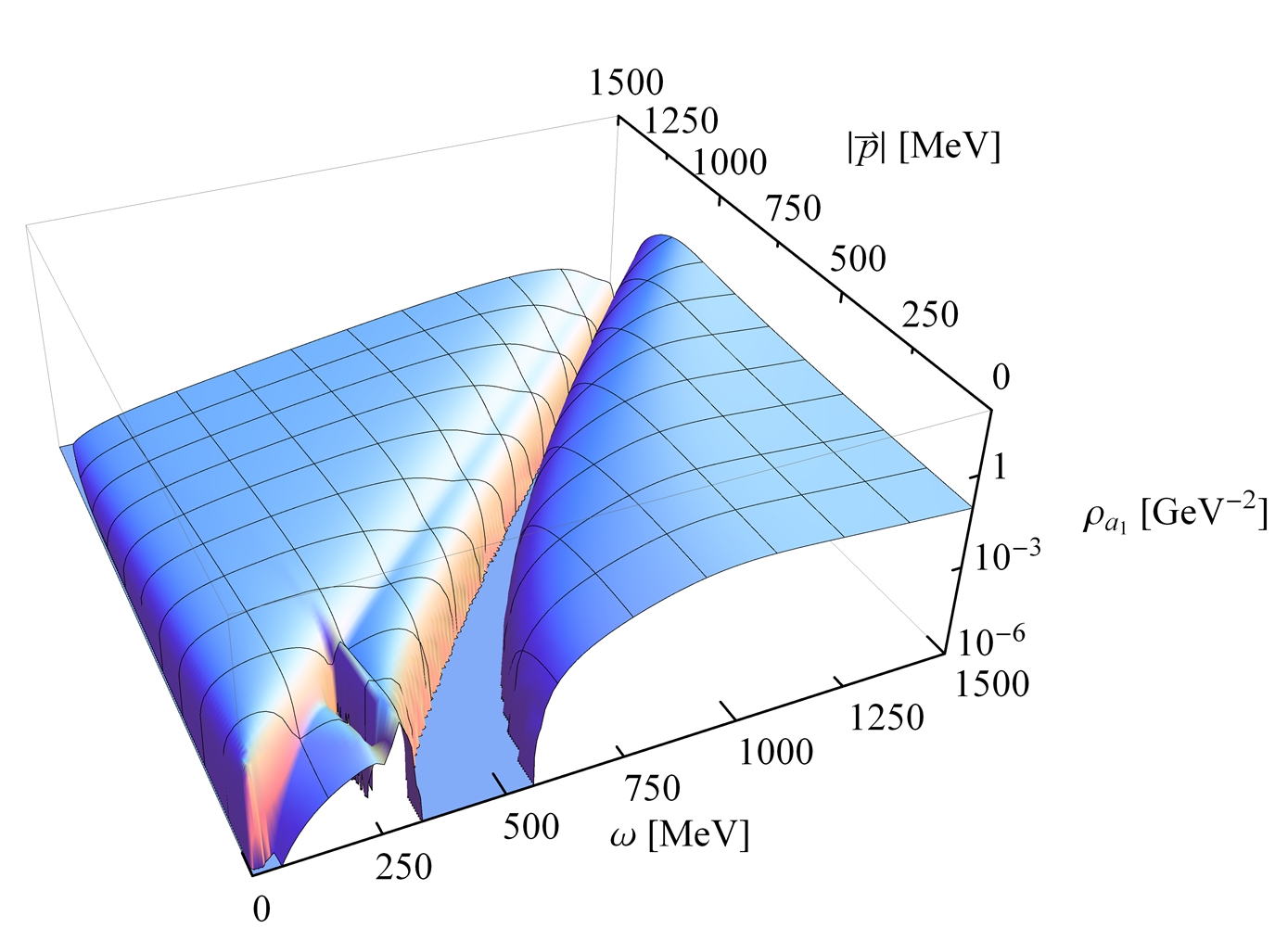}
	\caption{(color online) The transverse $\rho$ (left) and $a_1$ (right) spectral functions, $\rho_\rho (\omega, \vec{p})$ and $\rho_{a_1} (\omega, \vec{p})$, are shown versus external energy~$\omega$ and external spatial momentum $\vec{p}$ at $T=100$~MeV and $\mu=0$~MeV. The time-like regime, $\omega>|\vec{p}|$, is Lorentz-boosted to higher energies as the momentum increases, while the space-like regime, $\omega<|\vec{p}|$, is homogeneously filled up by the space-like processes.}
	\label{fig:spectral_3D}
\end{figure*}

At finite temperature the $a_1$ meson can capture a pion from the heat
bath to form a sigma meson, \mbox{$a_1^*+\pi\rightarrow \sigma$}. This
capture process can occur when $\omega = E_{k,\sigma} - E_{k,\pi}$, i.e.~for
the difference of the effective quasi-particle energies of $\sigma $
and $\pi$ at the momentum scale $k$, cf.~App.~\ref{sec:flow_equations}.   
It is therefore bounded by $\omega \leq m_{\sigma}-m_{\pi}$. When
the slopes of the quasi-particle branches of $\sigma $ and $\pi$ in the
scale $k$ get very close to one another during the FRG flow, i.e.~when
their difference $E_{k,\sigma} - E_{k,\pi}$ flows through a saddle point,
or an approximate one, the spectral density develops a peak analogous
to a van Hove singularity in the density of states in the electronic 
band-structure of solids. Such a van Hove peak is seen in the $a_1$
spectral function in the left column of
Fig.~\ref{fig:spectral_functions_temp_mu} for $T=100$~MeV and 150~MeV, 
i.e.~in the crossover region, just below the threshold of the capture
process at $\omega = m_{\sigma}-m_{\pi}$. 
As the difference between the sigma and the pion mass tends to zero,
this threshold moves to smaller and smaller energies when the
temperature is further increased. 

For temperatures around the crossover, the quark mass drops
significantly as well, leading to a shift of the associated threshold
to lower energies and a further broadening of the peaks in the
spectral functions. For very high temperatures pole and curvature
masses of the chiral partners degenerate, and the contribution from
the capture process disappears. The former behavior is due to the
direct link between the mass difference of $\rho$ and $a_1$ with the
chiral condensate, see \Eq{eq:masses}. The capture process disappears
for the same reason, namely because the $\pi$ and $\sigma$ masses
degenerate upon the melting of the condensate also. Hence, the
contributions of the dominant mesonic decay processes
\mbox{$\rho^*\rightarrow \pi+\pi$} 
and \mbox{$a_1^*\rightarrow \pi+\sigma$} to the respective spectral
functions degenerate as chiral symmetry gets restored as well. The
same holds for the $\pi$ and $\sigma$ tadpole contributions shown in
Fig.~\ref{fig:flow_equations_two-point}. Furthermore, the quarks
become the lightest degrees of freedom and thus give the dominant
contribution to the spectral functions in both channels. All these
effects together result in the complete degeneration of the spectral
functions of $\rho$ and $a_1$ and provide a direct connection to
chiral symmetry restoration.  

Coming now to the dependence on the chemical potential in the right column
of Fig.~\ref{fig:spectral_functions_temp_mu}, at a temperature of $T=
10$~MeV, we observe that both
spectral functions remain essentially unchanged from  $\mu = 0$ up to
values close to the critical endpoint,
reflecting the Silver Blaze property as already mentioned in
Sec.~\ref{subsec:masses}. Near the CEP, especially the $a_1$ spectral
function shows sensitive modifications which are mainly induced by the
dropping sigma mass. The main effect is that the threshold for the
process \mbox{$a_1^*\rightarrow \pi+\sigma$} moves to smaller energies
with another van Hove peak from an approximate saddle point forming in
$E_{k,\pi} + E_{k,\sigma}$ just above threshold. When hitting the CEP
exactly, which is difficult in a numerical calculation especially when
the mass of the $\sigma $ meson drops so suddenly close to the endpoint 
as in Fig.~\ref{fig:euclidean_masses_mu}, this threshold should be
located exactly at the pion mass since the sigma mass vanishes
completely there. The $\rho$ spectral function on the other hand shows only
small modifications. For very high chemical potentials and low
temperatures we again see a degeneracy of the $\rho$ and $a_1$
spectral functions. We note that $T=50$~MeV and $\mu=600$~MeV were
chosen for the last plot in Fig.~\ref{fig:spectral_functions_temp_mu}
in order to avoid the potentially problematic regime at low
temperatures and large chemical potentials, see the discussion in
Sec.~\ref{subsec:masses}.  

\subsection{Momentum dependence of spectral functions}\label{subsec:results_p}

As an instructive example for their momentum dependence we show the transverse $\rho$ and $a_1$ spectral functions at a temperature of $T=100$~MeV and $\mu=0$~MeV as a function of energy and momentum in Fig.~\ref{fig:spectral_3D}, see also App.~\ref{sec:flow_equations}.
 
In the case of the $\rho$ meson the time-like and space-like regimes are clearly separated. While the decay thresholds of the $\rho^*\rightarrow \pi + \pi$ and the $\rho^*\rightarrow \bar{\psi} + \psi$ process are Lorentz-boosted to higher energies as the spatial momentum increases, the space-like regime, where $\omega<|\vec{p}|$, is homogeneously filled up by the space-like processes, cf. Eq.~(\ref{eq:rho_spacelike}). We note that in the vacuum the spectral function would be zero in the space-like regime.

A similar behavior is observed for the $a_1$ spectral function. The
thresholds associated to the time-like processes are correctly boosted
to higher energies while the space-like regime does not show any
particular structure for the $a_1$ either. In contrast to the $\rho$
spectral function, however, there is no clear separation between the
time-like and space-like regime since the process $a_1^*+\pi
\rightarrow \sigma$ continuously connects both regions, see also
\cite{Tripolt2014a}. We also note that, unlike the threshold for this
capture process, the position of the van Hove peak in the
$a_1$ spectral function, which occurs close to this
threshold at $\omega=m_\sigma-m_\pi\approx 300$~MeV for $|\vec{p}|=0$,
remains essentially unaffected at finite momentum. Rather, this van
Hove peak eventually merges with the structureless space-like region.  

We conclude this section by noting that the momentum-dependent spectral functions are closely connected to the real-time propagators and thus allow also access to static properties of the medium, such as transport coefficients in the appropriate limits \cite{TripoltSmekalWambach2016a}.

\section{Summary}\label{sec:summary}

Within an effective description for low-energy QCD we have computed the in-medium spectral functions of the $\rho$ and $a_1$ mesons using the FRG method. This allows for a simultaneous description of the chiral phase structure of QCD and the behavior of the mesonic spectral functions at finite temperature and density within a unified framework. The main question we have addressed here was how the in-medium modifications of the spectral functions of the chiral partners $\rho$ and $a_1$ are connected to chiral symmetry restoration.

To illustrate this, we have used an effective low-energy model for two-flavor QCD based on a gauged linear sigma model augmented by dynamical constituent quarks. This construction was guided by requiring that both the dominant mesonic processes as well as the phase structure of QCD are treated on the same footing. Hence, in addition to the chiral partners, $\rho$ and $a_1$, we have included the lightest pseudoscalar and scalar mesons $\pi$ and $\sigma$ (the latter being identified with the $f_0(500)$ resonance). The momentum-independent self-interactions of the scalar and pseudoscalar mesons have been fully implemented by computing their effective potential. The leading-order interactions between the scalar and the vector sector have been included by invoking a gauge principle as in the gauged linear sigma model. The quark-meson interactions were constructed from QCD-inspired arguments concerning the dynamical generation of mesons from RG evolved quark-antiquark interactions in QCD. As we have demonstrated, this model allows for the simultaneous description of the chiral phase diagram in the ($T\!-\!\mu$)-plane and in-medium vector-meson spectral functions.

We have shown that by crossing the chiral phase boundary both at large temperature and large chemical potential, the spectral functions of $\rho$ and $a_1$ channel degenerate. The underlying mechanism which drives this degeneration becomes particularly lucid in the present approach. The mass splittings of the chiral partners decrease with the melting of the chiral condensate. As a result, the decay processes that determine the $\rho$ and $a_1$ spectral functions degenerate, which finally leads to degenerate spectral functions. Furthermore, the decreasing constituent quark mass towards chiral symmetry restoration triggers the melting of the resonance peaks in the spectral functions. We have demonstrated this behavior along the temperature and chemical potential directions of the phase diagram. As it turns out, the $\rho$ meson mass stays essentially constant while the $a_1$ shows a significant mass shift towards the $\rho$ when chiral symmetry restoration is approached. Chiral symmetry restoration is signaled by the degeneration of the spectral functions accompanied by a melting of the resonances rather than e.g.~a dropping of the mass as suggested by Brown and Rho \cite{Brown1991}. Hence, our findings are in favor of the``melting scenario" concerning the interpretation of experimental dilepton data. This is in line with the results from phenomenological models \cite{Rapp:1999ej} as well as results on the scaling of the vector meson masses in QCD \cite{Rennecke:2015eba}.

While our model captures the main qualitative features of the
vector-meson spectral functions, there are various directions left to
be explored in future studies in order to arrive at a complete
physical picture of the signatures of chiral symmetry restoration in
vector-meson spectral functions. The most obvious shortcoming of our
model is the lack of confinement. We show in
App.~\ref{sec:real_imag_part} that, while the qualitative features of
the spectral functions remain the same, the decay into on-shell quarks
give large contributions to the spectral functions deep in the
hadronic phase where confinement should prevent such processes. In the
present context, confinement manifests itself in positivity violation
of the quark spectral functions, see e.g.~\cite{Qin:2013ufa}, and this
should be accounted for in a more realistic description of chiral
symmetry restoration. The incorporation of gluon fluctuations, which
will certainly become relevant in the vicinity of the phase transition
and at large frequencies/momenta, is also a crucial step in this
direction and may be incorporated along the lines of
\cite{Rennecke:2015eba}. This would also make the necessity of model
parameter tuning obsolete. In order to capture the full momentum
dependence of the propagators, we need to go beyond the classical
momentum dependence considered here. This can be done e.g.~by
including non-trivial wavefunction renormalizations in the gradient
expansion of the effective action. It has been shown in
\cite{Helmboldt:2014iya}, that this reduces the difference between
curvature and pole masses, especially for the light
mesons. Phenomenologically viable    
spectral functions also require the inclusion of the feedback of the
vector mesons themselves as well as baryonic degrees of freedom, both
of which have been neglected here. Especially given the fact that our
approach allows for the computation of spectral functions also at
arbitrarily large chemical potential calls for an extension to include 
the dynamics of baryons and nuclear-matter effects in a chiral
effective theory that is capable of describing both, the liquid-gas
transition of nuclear matter and the chiral transition in a unified
framework \cite{Weyrich:2015hha}. Several of these issues have been
addressed already within the FRG, showing that this method is flexible
enough to tackle these ambitious tasks in the future. 


\acknowledgments 
This work was supported by the German Federal Ministry of Education
and Research (BMBF), grant no. 05P16RDFC1. FR was supported by the FWF grant P24780-N27 and the DFG via SFB 1225 (ISOQUANT).



\appendix

\section{Derivation of flow equations}
\label{sec:flow_equations}
In the following we provide further details on our theoretical setup
as well as explicit expressions for the flow equations, see also
\cite{Tripolt2014, Tripolt2014a}. The effective potential only
contains the scalar meson as well as quark-antiquark fluctuations and
is hence given by solving the flow equation obtained by applying the
Wetterich equation, Eq.~(\ref{eq:wetterich}), to the ansatz for the
effective action of the quark-meson model as done in
\cite{Tripolt2014, Tripolt2014a,Schaefer:2004en}, which yields 
\begin{equation}
\label{eq:flow_eq_eff_pot}
\begin{split}
\partial_k U_k =& \frac{k^4}{12 \pi^2}\Bigg[
\frac{1}{E_{k,\sigma}} \coth{\left(\frac{E_{k,\sigma}}{2T}\right)} +\frac{3}{E_{k,\pi}} \coth{\left(\frac{E_{k,\pi}}{2T}\right)}\\
&-\frac{2N_fN_c}{E_{k,\psi}} \Bigg(\tanh{\left(\frac{E_{k,\psi}-\mu}{2T}\right)}\\
&+\tanh{\left(\frac{E_{k,\psi}+\mu}{2T}\right)}\Bigg)\Bigg].
\end{split}
\end{equation}
The effective quasi-particle energies are given by
\begin{equation}
\label{eq:energies}
E_{k,\alpha}\equiv\sqrt{k^2+m_{k,\alpha}^2}, \qquad \alpha \in
\{\pi,\sigma,\rho,a_1,\psi\}\,, 
\end{equation}
with effective masses of the quarks and mesons
\begin{alignat}{2}
\label{eq:masses}
&m_{k,\pi}^2&&=2U_k',\\
&m_{k,\sigma}^2&&=2U_k'+4 U_k''\phi_0^2,\\
&m_{k,\rho}^2&&=m_{k,V}^2,\\
&m_{k,a_1}^2&&=m_{k,V}^2 + g^2 \phi_0^2,\\
&m_{k,\psi}^2&&=h_S^2\phi_0^2,
\end{alignat}
where primes denote derivatives with respect to the chiral invariant $\phi^2 \equiv \sigma^2+\vec{\pi}^2$ and $\phi_0^2=\sigma_0^2$ is meant to be the global minimum.

Flow equations for two-point functions of vector mesons transversal and longitudinal to the heat bath are defined as
\begin{align}
\label{eq:projection_flow_eq_two-point_functions}
\begin{split}
\partial_k\Gamma_{k,\rho/a_1}^{(2),\perp}(p) &= \frac{1}{2(N_f^2-1)}\,\text{Tr}\,\Bigl[\Pi^{T,\perp}_{\mu\nu}(p) \partial_k \left(\Gamma_{k,\rho/a_1}^{(2)}(p)\right)_{\nu\sigma}^{fg} \Bigr]\,,\\
\partial_k\Gamma_{k,\rho/a_1}^{(2),\parallel}(p) &= \frac{1}{(N_f^2-1)}\, \text{Tr}\,\Bigl[\Pi^{T,\parallel}_{\mu\nu}(p) \partial_k \left(\Gamma_{k,\rho/a_1}^{(2)}(p)\right)_{\nu\sigma}^{fg} \Bigr]\,,
\end{split}
\end{align}
where $f,g \in \{1,..,(N_f^2-1)\}$ are adjoint flavor indices. The
three dimensionally transverse and four dimensionally longitudinal
projection operators are defined by
\begin{align}
\label{eq:projection_heat-bath}
\begin{split}
\Pi^{T,\perp}_{\mu\nu}(p) &= \begin{cases} 0 & \text{if}\; \mu=0\; \text{or}\; \nu=0\\ \delta_{\mu\nu}-\frac{p_\mu p_\nu}{\vec{p}^2} & \text{else} \end{cases}\,,\\
\Pi^{T,\parallel}_{\mu\nu}(p) &= \delta_{\mu\nu}-\frac{p_\mu p_\nu}{p^2} - \Pi^T_{\mu\nu}(p)\,,
\end{split}
\end{align}
with
\begin{equation}
\Pi^{T}_{\mu\nu}(p) = \Pi^{T,\perp}_{\mu\nu}(p)+\Pi^{T,\parallel}_{\mu\nu}(p) = \delta_{\mu\nu}-\frac{p_{\mu}p_\nu}{p^2}.
\end{equation}
For vanishing external spatial momentum, $\vec{p} = 0$, we find that
longitudinal and transverse parts of the flow equations coincide
\begin{equation}
\partial_k\Gamma_{k,\rho/a_1}^{(2),\perp}(p_0)=\partial_k\Gamma_{k,\rho/a_1}^{(2),\parallel}(p_0).
\end{equation}

As illustrated diagrammatically in Fig.~\ref{fig:flow_equations_two-point} the flow equations for the $\rho$ and $a_1$ two-point functions read more explicitly
\begin{align}
\label{eq:explicit_flow_eq_two-point_functions}
\partial_k\Gamma_{k,\rho}^{(2),\perp}(p_0) &= J^{\pi\pi}_{k,\rho}(p_0)-\frac{1}{2}I^{\pi}_{k,\rho}-2 J^{\psi\bar{\psi}}_{k,\rho}(p_0),\\
\partial_k\Gamma_{k,a_1}^{(2),\perp}(p_0) &= J^{\sigma\pi}_{k,a_1}(p_0)+J^{\pi\sigma}_{k,a_1}(p_0)\nonumber\\
&\quad-\frac{1}{2}I^{\pi}_{k,a_1}-\frac{1}{2}I^{\sigma}_{k,a_1}-2 J^{\psi\bar{\psi}}_{k,a_1}(p_0).
\end{align}
The loop functions are defined as
\begin{align}
\label{eq:loop_functions}
I^{\beta}_{k,\alpha} &\equiv \text{Tr}\Big[\partial R_{k}^{\beta}(q) G_{k,\beta}(q)\Gamma^{(4)}_{k,\beta\alpha}G_{k,\beta}(q)\Big],\\
J^{\beta\gamma}_{k,\alpha}(p)&\equiv\text{Tr}\Big[  \partial R_{k}^{\beta}(q+p) G_{k,\beta}(q+p)\Gamma^{(3)}_{k,\beta\gamma\alpha}\nonumber\\
&\qquad G_{k,\gamma}(q)\Gamma^{(3)}_{k,\gamma\beta\alpha}G_{k,\beta}(q+p)\Big],
\end{align}
with scale-dependent regulated propagator
\begin{equation}
\label{eq:propagator}
G_{k,\alpha}(q) \equiv \left(\Gamma^{(2)}_{k,\alpha}+R_{k}^{\beta}(q)\right)^{-1}.
\end{equation}
The trace includes summations over all internal indices as well as a loop momentum integration which, for finite temperature, turns into a spatial integration and a summation over Matsubara modes.

The regulator function has to be chosen appropriately for the different types of fields \cite{Pawlowski:2015mlf}. In this work we use the following three-dimensional regulator functions
\begin{alignat}{2}
&R_k^{\sigma/\pi}(q) &&= (k^2-\vec{q\,}^2) \theta (k^2-\vec{\,q}^2 ),\label{eqs:regulator_scalar}\\
&R_k^{\rho/a_1}(q) &&=\Pi^{T}_{\mu\nu}(q) (k^2-\vec{q\,}^2 ) \theta (k^2-\vec{\,q}^2 ),\label{eqs:regulator_vector}\\
&R_k^{\psi}(q)  &&= \mathrm{i} \slashed{\vec q} (\sqrt{k^2/\vec q^{\,2}}-1)
\theta \left(k^2-\vec q^{\,2} \right)\label{eqs:regulator_quarks}.
\end{alignat}
Explicit expressions for the 3- and 4-point vertices, $\Gamma_{k}^{(3)}$ and $\Gamma_{k}^{(4)}$, are given by
\begin{alignat}{2}
\label{eq:vertices}
&\Gamma_{k,\bar \psi \psi \rho_i^{\mu}}^{(3)}&&=\mathrm{i} h_V \gamma^{\mu}\tau_i,\\
&\Gamma_{k,\bar \psi \psi a_{1,i}^{\mu}}^{(3)}&&=\mathrm{i} h_V \gamma^{\mu}\gamma^5\tau_i,\\
&\Gamma_{k,\pi_k\pi_j\rho_i^{\mu}}^{(3)}&&=\mathrm{i}g\epsilon_{ijk}(q_{\pi_k}^{\mu}-q_{\pi_j}^{\mu})\left(1-\frac{g^2 \sigma_0}{m_{k,a_1}^2}\right), \\
&\Gamma_{k,\sigma\pi_ja_{1,i}^{\mu}}^{(3)}&&=\mathrm{i}g\delta_{jl}(q_{\sigma}^{\mu}-q_{\pi_j}^{\mu}), \\
&\Gamma_{k,\pi_l\pi_k\rho_j^{\nu}\rho_i^{\mu}}^{(4)}&&=g^2\delta^{\mu\nu}\left(2\delta_{ij}\delta_{kl}-\delta_{ik}\delta_{jl}-\delta_{il}\delta_{jk}\right),\\
&\Gamma_{k,\sigma\sigma a_{1,j}^{\nu}a_{1,i}^{\mu}}^{(4)}&&=2g^2\delta_{ij}\delta^{\mu\nu},\\
&\Gamma_{k,\pi_l\pi_k a_{1,j}^{\nu}a_{1,i}^{\mu}}^{(4)}&&=g^2\delta^{\mu\nu}\left(\delta_{ik}\delta_{jl}+\delta_{il}\delta_{jk}\right).
\end{alignat}

The flow of the scale-dependent coupling $m_{k,V}^2$ is defined by projecting on the $\rho$ mass in following way
\begin{align}
\label{eq:flow_mV}
\begin{split}
\partial_k m_{k,V}^2 &= \frac{1}{2(N_f^2-1)} \lim_{p \rightarrow 0} \text{Tr}\!\left.\left(\!\Pi_{\mu\nu}^{T,\perp} \frac{\delta^2\partial_k \Gamma_k}{\delta \rho_i^\mu \delta\rho_j^\nu} \right)\right|_{\phi=\phi_0}\\
&= \partial_k m_{k,\rho}^2\,.
\end{split}
\end{align}
We note that in our approach the Euclidean curvature mass of the $\rho$ meson is given by the vector meson coupling $m_{k,V}$, thus their flow equations coincide.

\section{Analytic continuation and spectral functions}
\label{sec:analytic_continuation}

\begin{figure*}[t!]
	\includegraphics[width=0.49\textwidth]{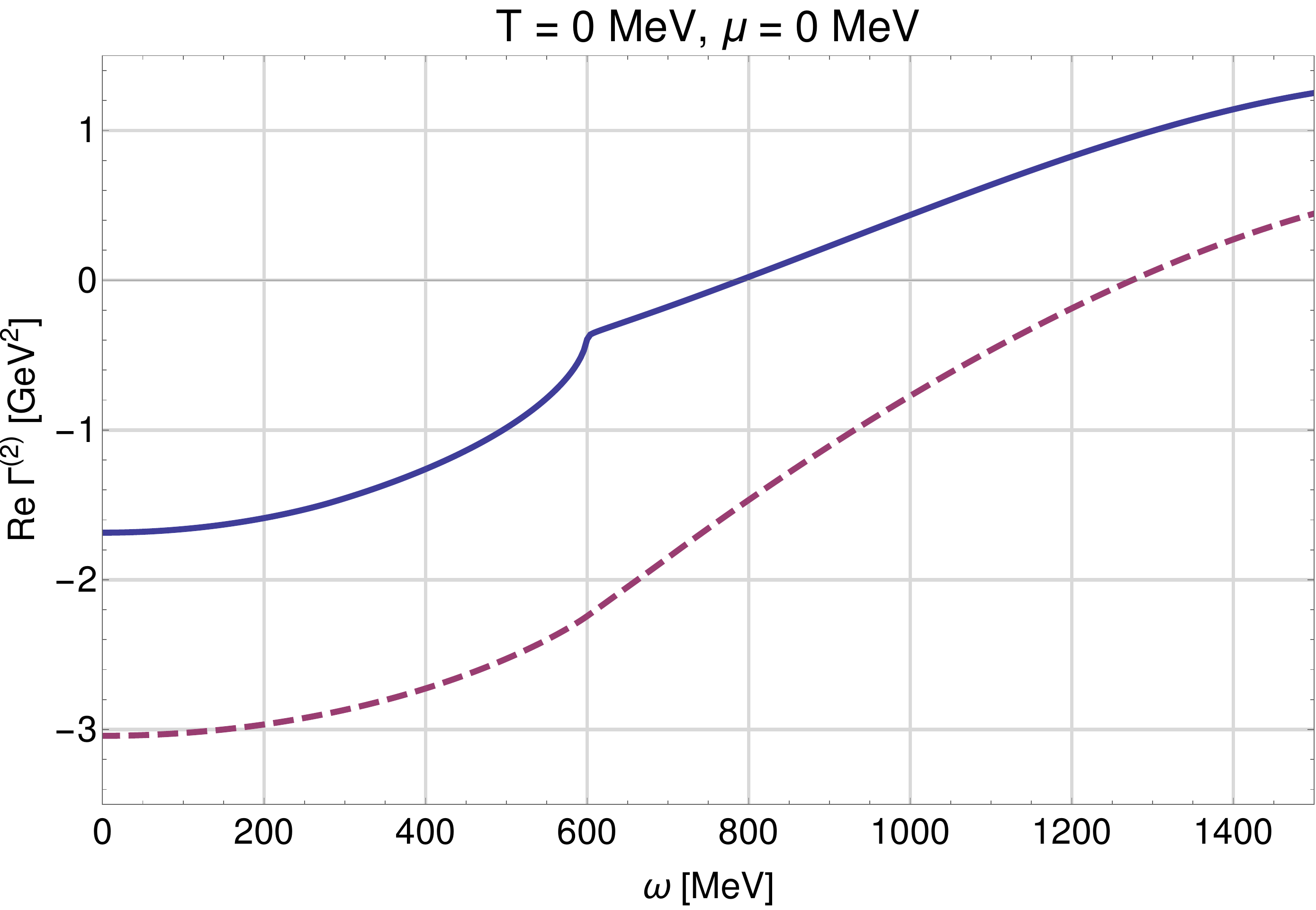}
	\includegraphics[width=0.49\textwidth]{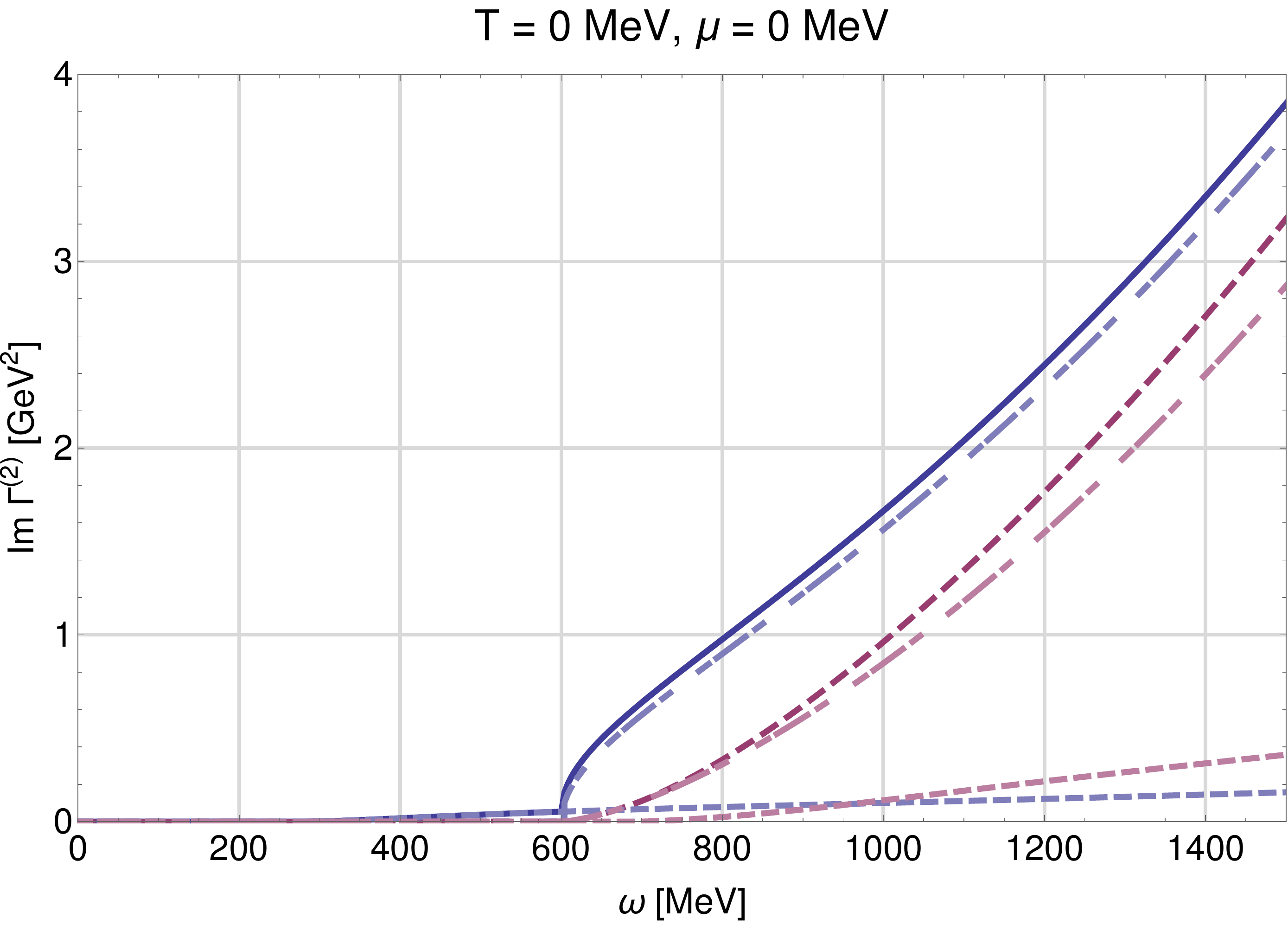}
	\caption{Real (left) and imaginary (right) part of the retarded two-point function of the $\rho$ (blue) and $a_1$ (dashed red) meson vs. external energy $\omega$ at $T=0$ MeV and $\mu = 0$ MeV. The different contributions to the imaginary parts are shown in light dashed lines (right): \mbox{$\rho^*/a_1^*\rightarrow \psi+\bar{\psi}$} (upper dashed lines) and \mbox{$\rho^*\rightarrow \pi+\pi$} for the $\rho$ meson and \mbox{$a_1^*\rightarrow \pi+\sigma$} for the $a_1$ meson (lower dashed lines).}\label{fig:real_imag}
\end{figure*}

In order to obtain flow equations for real-time (retarded) two-point functions an analytic continuation from imaginary to real energies has to be performed. We use the following two-step analytic continuation procedure which was developed in \cite{Tripolt2014, Tripolt2014a} for the FRG. In a first step the periodicity of the bosonic and fermionic occupation numbers, which result from the Matsubara summation over the loop energy, w.r.t.~the discrete external  Euclidean energy $p_0$ is exploited, i.e. 
\begin{equation}
\label{eq:continuation1}
n_{B,F}(E+\I p_0)\rightarrow n_{B,F}(E).
\end{equation}
In a second step, the Euclidean energy $p_0$ is replaced by a continuous real frequency $\omega$ in the following way,
\begin{equation}
\label{eq:continuation2}
\Gamma^{(2),R}(\omega,\vec p)=-\lim_{\epsilon\to 0} \Gamma^{(2),E}(p_0=-\I(\omega+\I\epsilon), \vec p),
\end{equation}
where the limit $\epsilon\to 0$ can be taken exactly for the imaginary part of the two-point functions, see App.~\ref{sec:eps0}, while for the real part we use a small value of $\epsilon=0.1$~MeV or $\epsilon=1$~MeV in our numerical implementation. 

The flow equations for the retarded two-point functions are then solved using the grid method with the initial values given by
\begin{eqnarray}
\label{eq:UV_rho} 
\Gamma^{(2),R}_{\Lambda,\rho}(\omega, \vec{p})&=&(\omega+\I\epsilon)^2-\vec{p}^{\,2}-m^2_{\Lambda,\rho}, \\
\label{eq:UV_a1} 
\Gamma^{(2),R}_{\Lambda,a_1}(\omega, \vec{p})&=&(\omega+\I\epsilon)^2-\vec{p}^{\,2}-m^2_{\Lambda,a_1}.
\end{eqnarray}

The spectral functions are essentially given by the imaginary part of the retarded propagator,
\begin{equation}
\rho(\omega,\vec p)=-\frac{1}{\pi}\text{Im} G^R(\omega,\vec p),
\end{equation}
which can be expressed in terms of the retarded two-point function as
\begin{equation}
\rho(\omega,\vec p)=\frac{1}{\pi}\frac{\text{Im}\,\Gamma^{(2),R}(\omega,\vec p)}{\left(\text{Re}\,
	\Gamma^{(2),R}(\omega,\vec p)\right)^2+\left(\text{Im}\,\Gamma^{(2),R}(\omega,\vec p)\right)^2}.
\end{equation}

\section{Analytic imaginary parts}
\label{sec:eps0}
The limit $\epsilon\rightarrow 0$ in the definition of the retarded
two-point functions, Eq.~(\ref{eq:continuation2}), can be performed
analytically for the imaginary part of the two-point functions in the
following way. For simplicity, we discuss here the case of vanishing
external spatial momentum. First we note that the imaginary parts of
the retarded two-point functions are obtained from those of 
the prefactors of the loop functions in App.~\ref{sec:loop_functions}
with Dirac-Sokhotsky identities,
\begin{align}
\lim_{\epsilon \rightarrow 0}\text{Im}\left(\frac{1}{\omega+ \I\epsilon -E_\alpha\pm E_\beta}\right)\rightarrow -\pi\delta(\omega-E_\alpha \pm E_\beta),\nonumber\\
\lim_{\epsilon \rightarrow
  0}\text{Im}\left(\frac{1}{(\omega+\I\epsilon -E_\alpha\pm E_\beta)^2}\right)\rightarrow \pi\delta'(\omega-E_\alpha\pm E_\beta). \nonumber 
\end{align}
Expressing these delta functions in terms of $k$ the flow equations
for the imaginary parts of the retarded 2-point functions then collapse
to the $k=k_0$ values which solve
\begin{align}
\omega-E_{k,\alpha} \pm E_{k,\beta} = 0,
\end{align}
so that the integrated flow equations reduce to the following form,
\begin{align}
\int_{k_{\text{UV}}}^{k_{\text{IR}}}dk\:\partial_k \text{Im} \Gamma_k^{(2)}
=&\int_{k_{\text{UV}}}^{k_{\text{IR}}}dk\: \big(f(k) \delta(k-k_0) \nonumber\\
& \hskip 2cm + g(k) \delta'(k-k_0) \big)\nonumber\\
=& -f(k_0)+g'(k_0). 
\end{align}
Where the generic functions $f(k)$ and $g(k)$ contain the
derivatives $E'_{k,\alpha} \pm E'_{k,\beta}$ of the quasi-particle
energies w.r.t.~the momentum scale $k$ in the denominator 
which gives rise to the van Hove singularities at saddle points.  

An analogous procedure is possible at finite external spatial momentum.

\section{Real and imaginary part of retarded two-point functions}
\label{sec:real_imag_part}

As described above, real and the imaginary parts of the flow equations
for the vector meson retarded two-point functions are solved
separately. In the following we present results for these parts and
their compositions. The zero crossings of the real parts are here used
to define the mass of the particle. It agrees with the physical pole mass of a
stable particle, if the imaginary part is zero at this energy as
well. Otherwise it locates, at least approximately,  the peak of a
resonance whose width is determined by the imaginary part of the
corresponding zero of the retarded two-point function 
on the unphysical Riemann sheet.

In Fig.~\ref{fig:real_imag} the real (left) and the imaginary parts
(right) of the vacuum $\rho$ and $a_1$ retarded two-point functions
are shown. The real part of the $\rho$ retarded two-point function
shows a zero-crossing at \mbox{$\omega = 789.3$ MeV}. The related
imaginary part begins to raise significantly when the quark-antiquark
decay becomes possible at around \mbox{$\omega \approx 600$ MeV}, the
$\rho$ peak is hence strongly suppressed. For the $a_1$ meson, this
suppression is even stronger since the real part has a zero-crossing
at \mbox{$\omega = 1274.7$ MeV} where the imaginary part already assumes
large values.  

In order to see how strong the particular processes contribute to the
imaginary part, these contributions are plotted separately on the
right in Fig.~\ref{fig:real_imag}. We see that for the $\rho$ as well
as for the $a_1$ imaginary part the quark-antiquark decay process
dominates clearly over the mesonic decay processes, even in the
vacuum, where physically this decay should not be possible. However,
in our approach this process is naturally present since we have no
mechanism which describes confinement and it therefore needs further
investigations how it can be suppressed or even removed in a
physically reasonable way of modeling confinement.  

\begin{figure*}[t!]
	\includegraphics[width=0.49\textwidth]{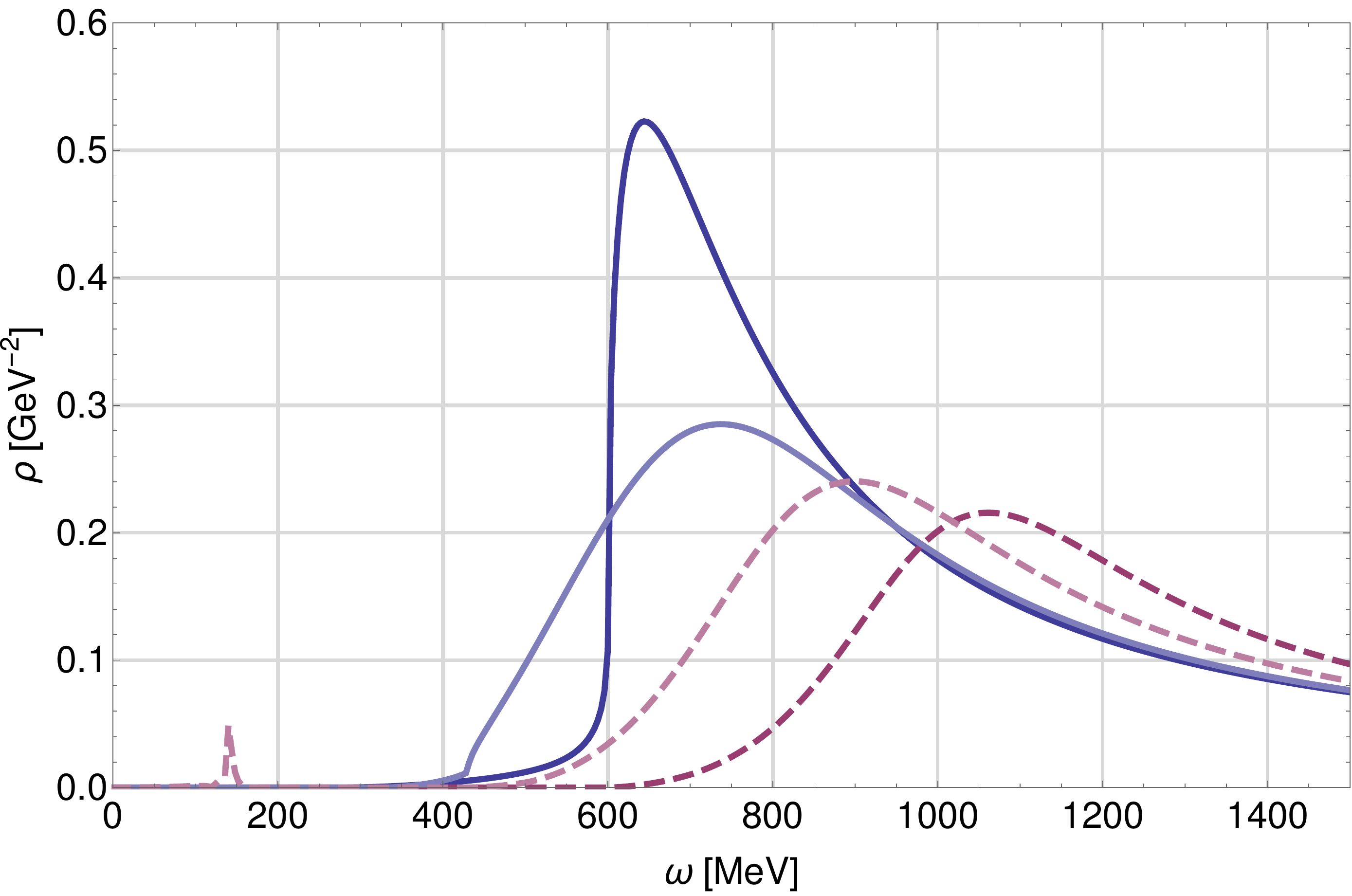}
	\includegraphics[width=0.49\textwidth]{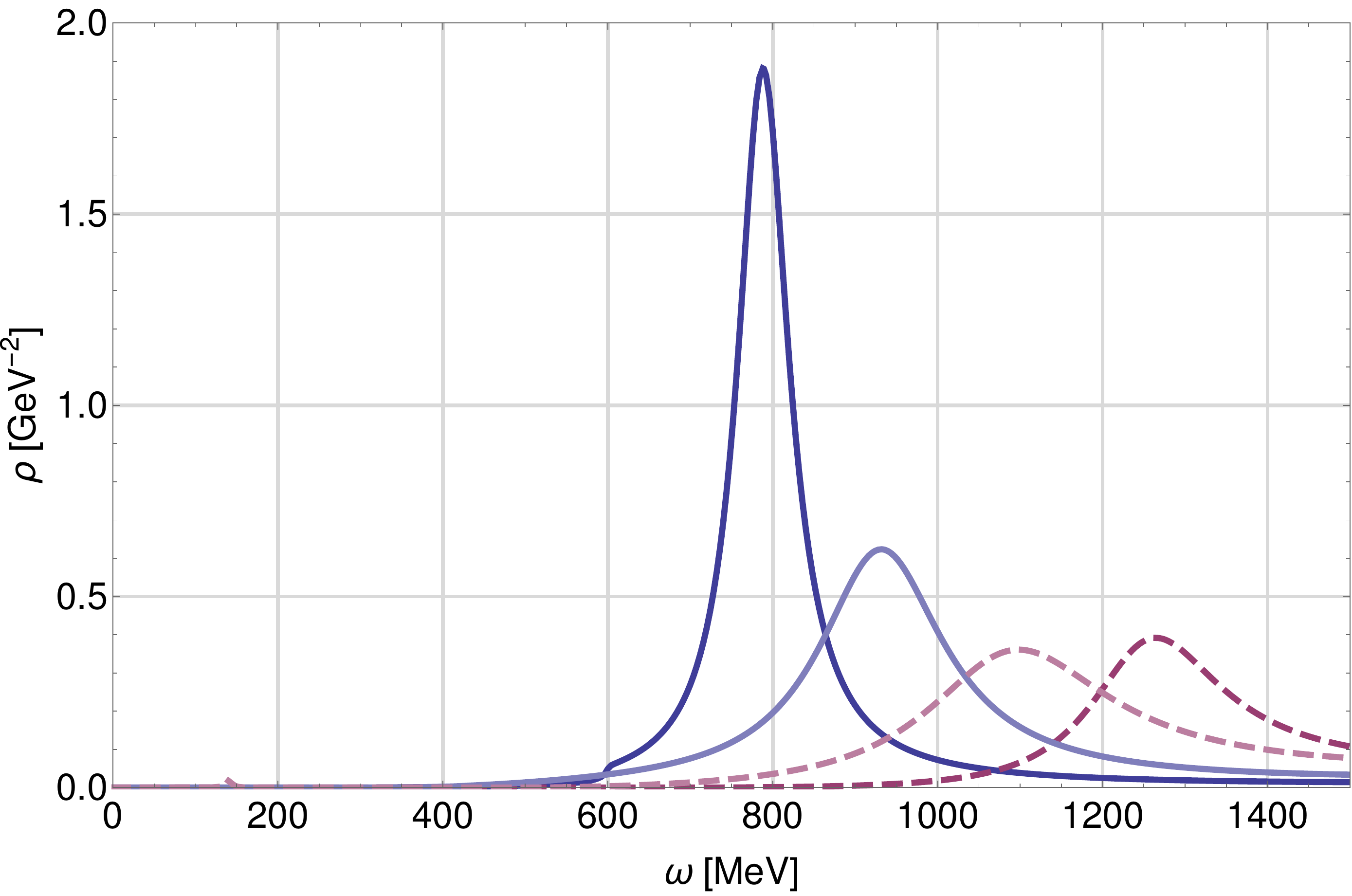}
	\caption{Spectral functions of the $\rho$ (blue) and $a_1$ (dashed red) meson at $T=0$ MeV (normal color) and $T=150$ MeV (light color) both for vanishing chemical potential in linear scales with (left) and without (right) quark-antiquark decay channel.}\label{fig:linear_plots}
\end{figure*}

\begin{figure}[h]
	\includegraphics[width=0.49\textwidth]{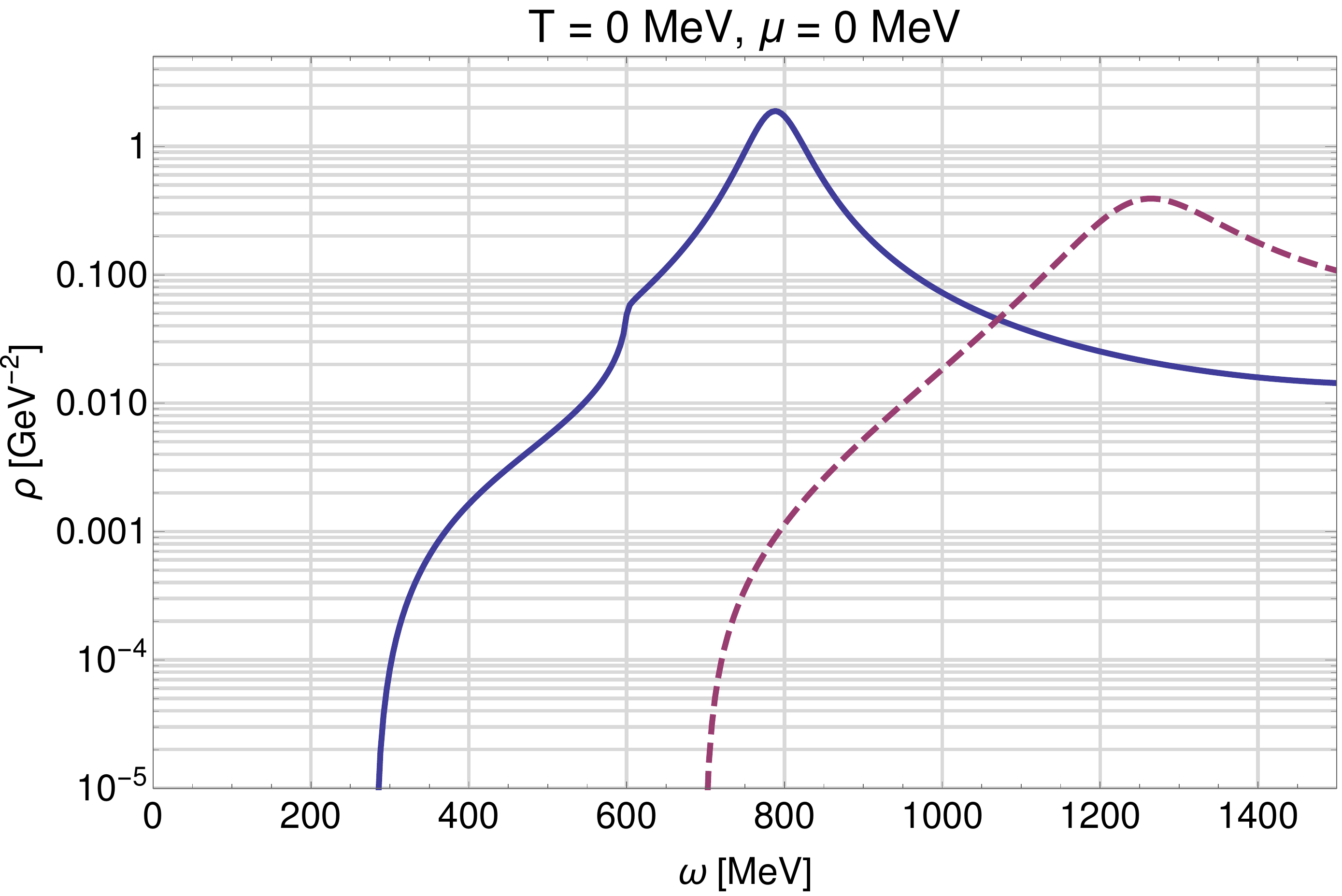}
	\caption{Vacuum spectral functions of the $\rho$ (blue) and $a_1$ (dashed red) meson without quark-antiquark decay channel.}\label{fig:spectral_functions_quarks}
\end{figure}

To isolate the effects of the quark-antiquark decay channel in the
spectral functions, \Fig{fig:linear_plots} shows the full $\rho$ and
$a_1$ spectral functions (left) in comparison with spectral functions where the
quark-antiquark contributions have been removed by hand from the
imaginary parts of the two-point functions (right) in the vacuum and
at $T=150$ MeV for vanishing chemical potential, here both plotted on a
linear scale. The increasing temperature leads to a melting especially
of the $\rho$ meson spectral function and a shift of the pole mass peaks towards
each other in both cases. Additionally, due to the thermal capture
process, in the $a_1$ spectral function a small peak arises for lower
energies. Without the quark-antiquark decay channel, the peaks are
less suppressed and concentrated around the masses defined by the
zero-crossing of the real part. As shown in \Fig{fig:real_imag}
on the right, the quarks above threshold
give the by far dominant contributions to the imaginary
parts and therefore shift the positions of the mass peaks in the
spectral functions as well, which can be seen by comparing both sides in
\Fig{fig:linear_plots}. For a better comparison with
\Fig{fig:spectral_functions_temp_mu} in the main text, the
vacuum spectral functions without the quark-antiquark decay channel
are also plotted on logarithmic scales in
\Fig{fig:spectral_functions_quarks}. Here we see a broad 
peaks concentrated around the pole masses of the $\rho$ and the $a_1$ meson.

\section{Available processes}\label{sec:processes}
\begin{figure*}[t]	
	\includegraphics[width=0.32\columnwidth]{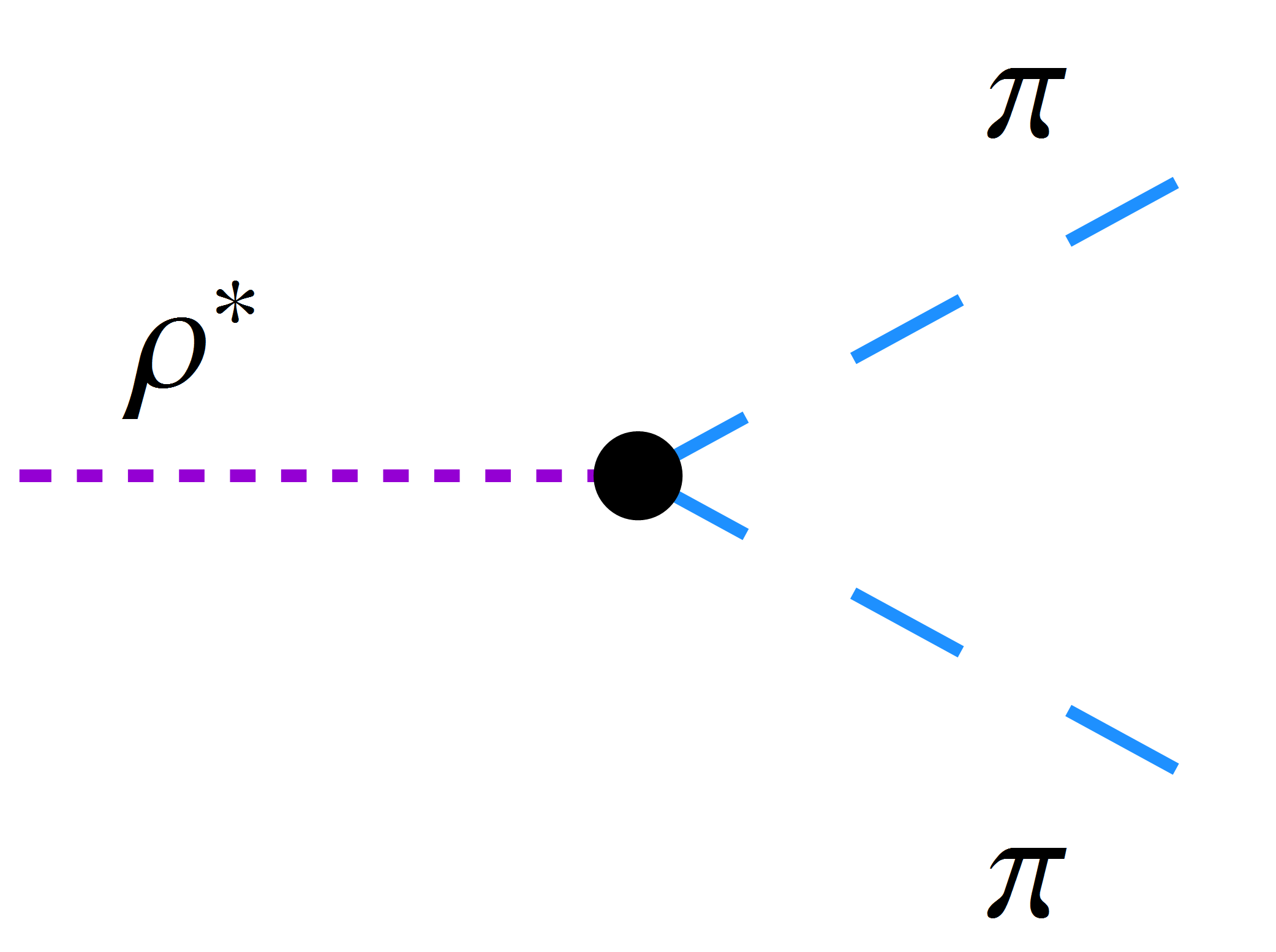}\hspace{15mm}
	\includegraphics[width=0.32\columnwidth]{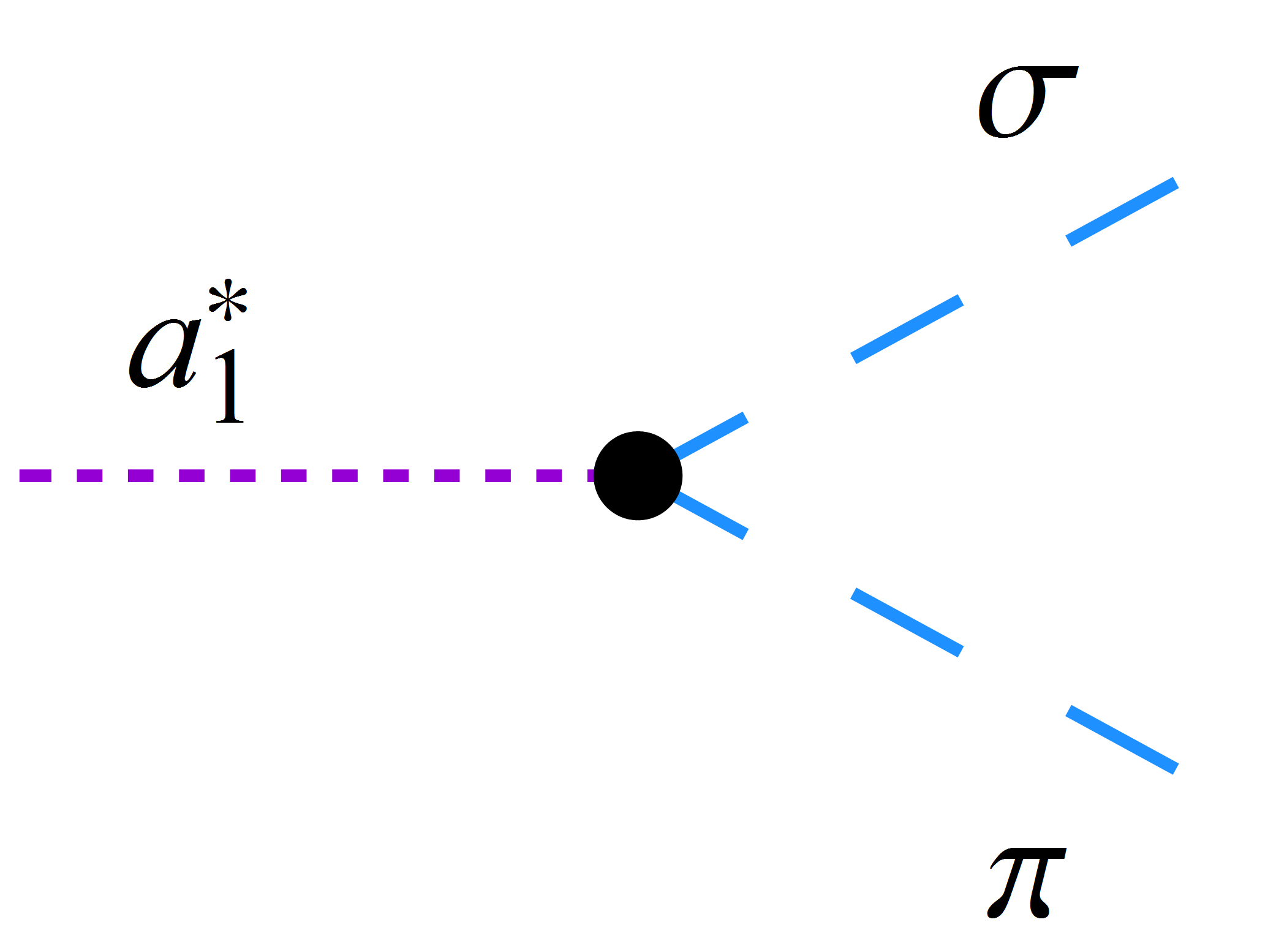}\hspace{15mm}
	\includegraphics[width=0.32\columnwidth]{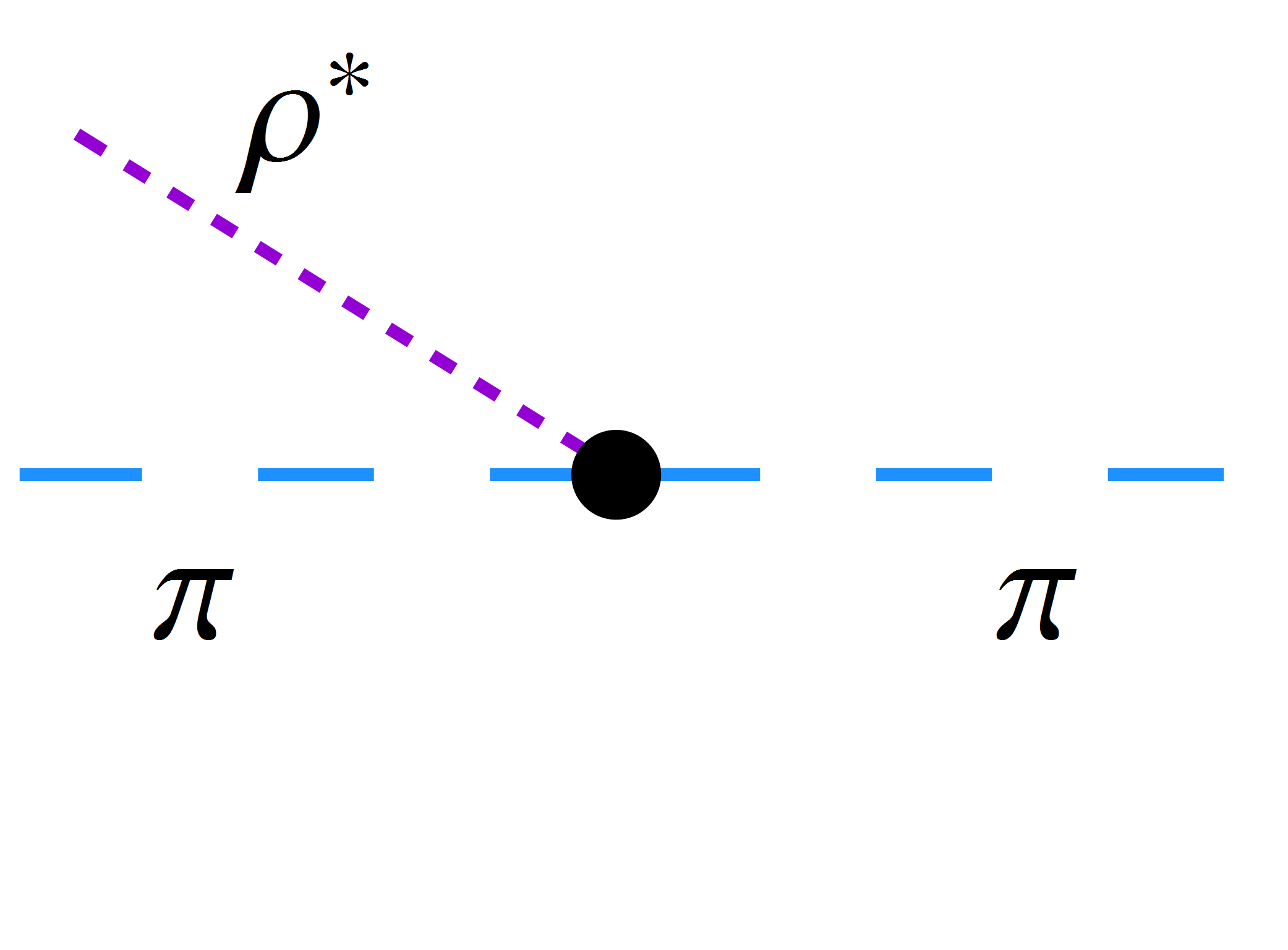}\hspace{15mm}
	\includegraphics[width=0.32\columnwidth]{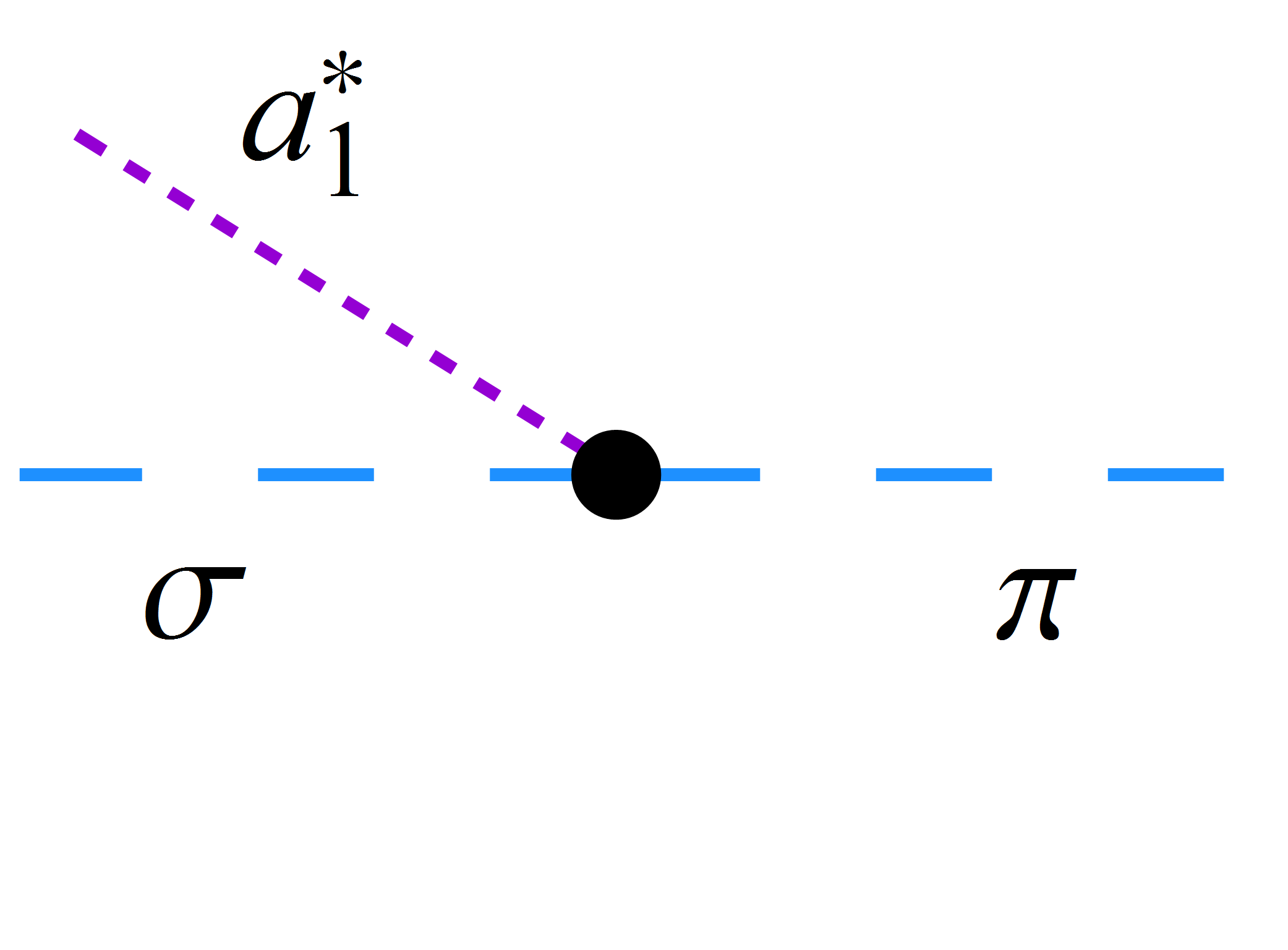}\\[3mm]
	\includegraphics[width=0.32\columnwidth]{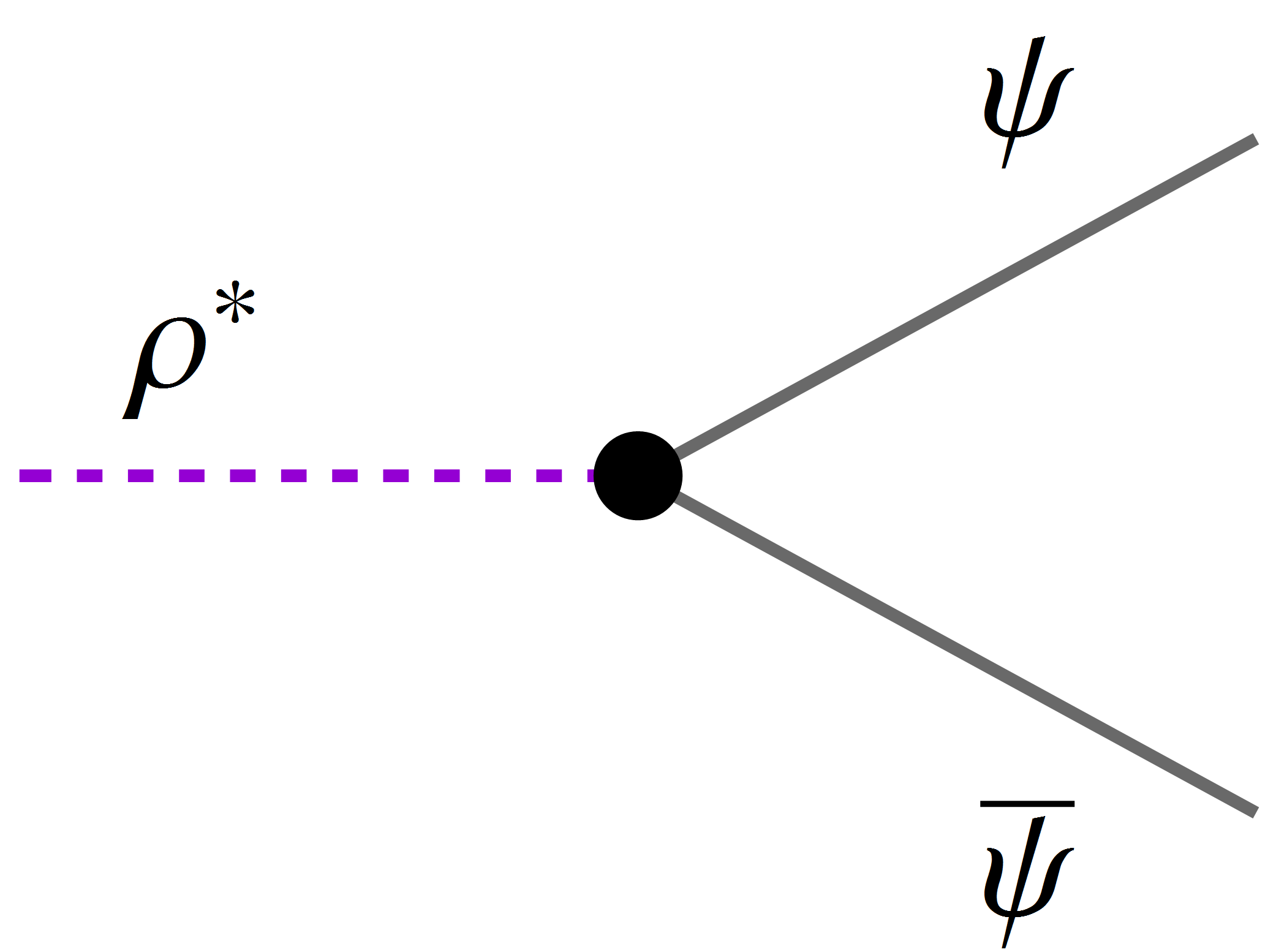}\hspace{15mm}
	\includegraphics[width=0.32\columnwidth]{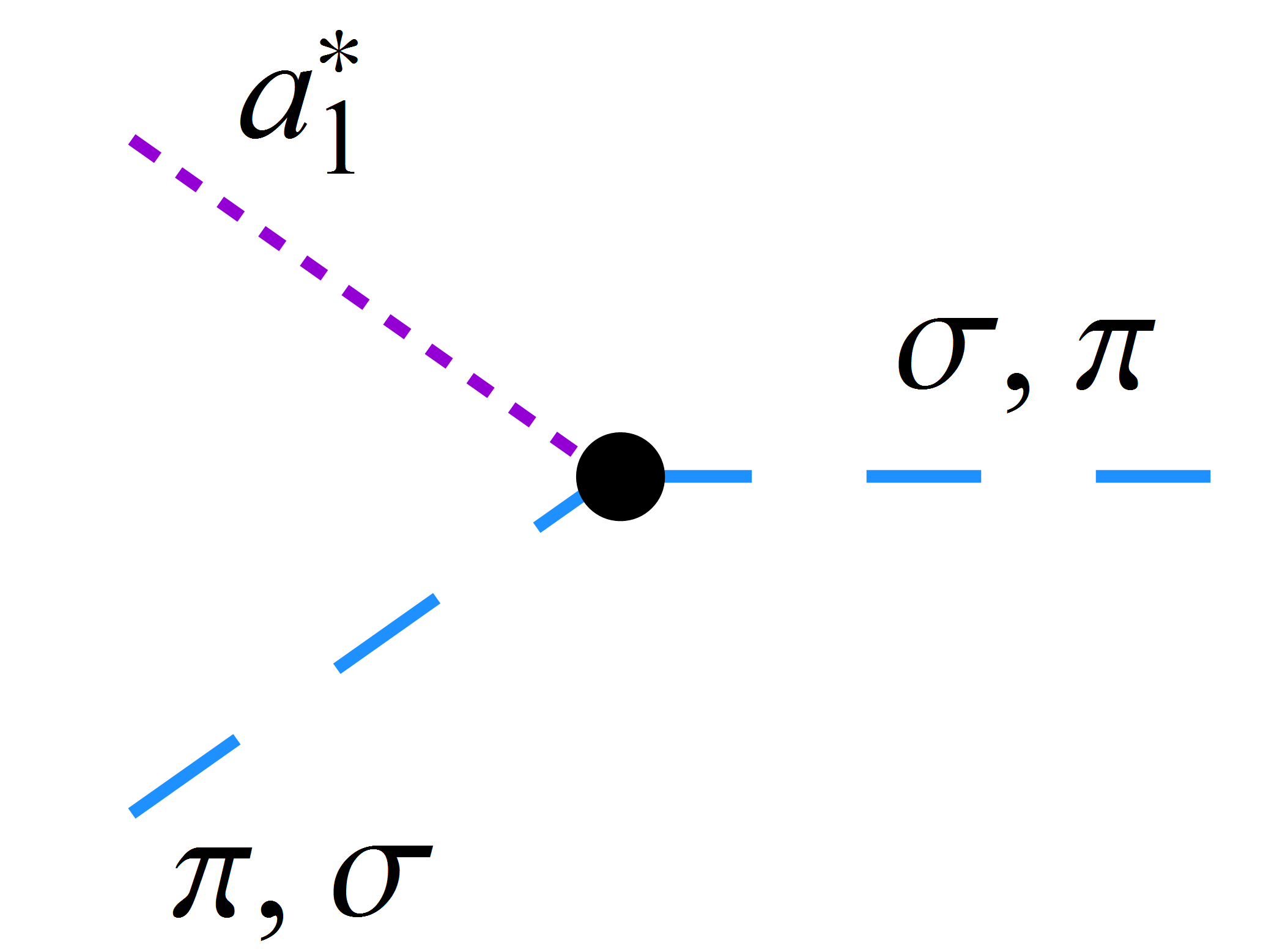}\hspace{15mm}
	\includegraphics[width=0.32\columnwidth]{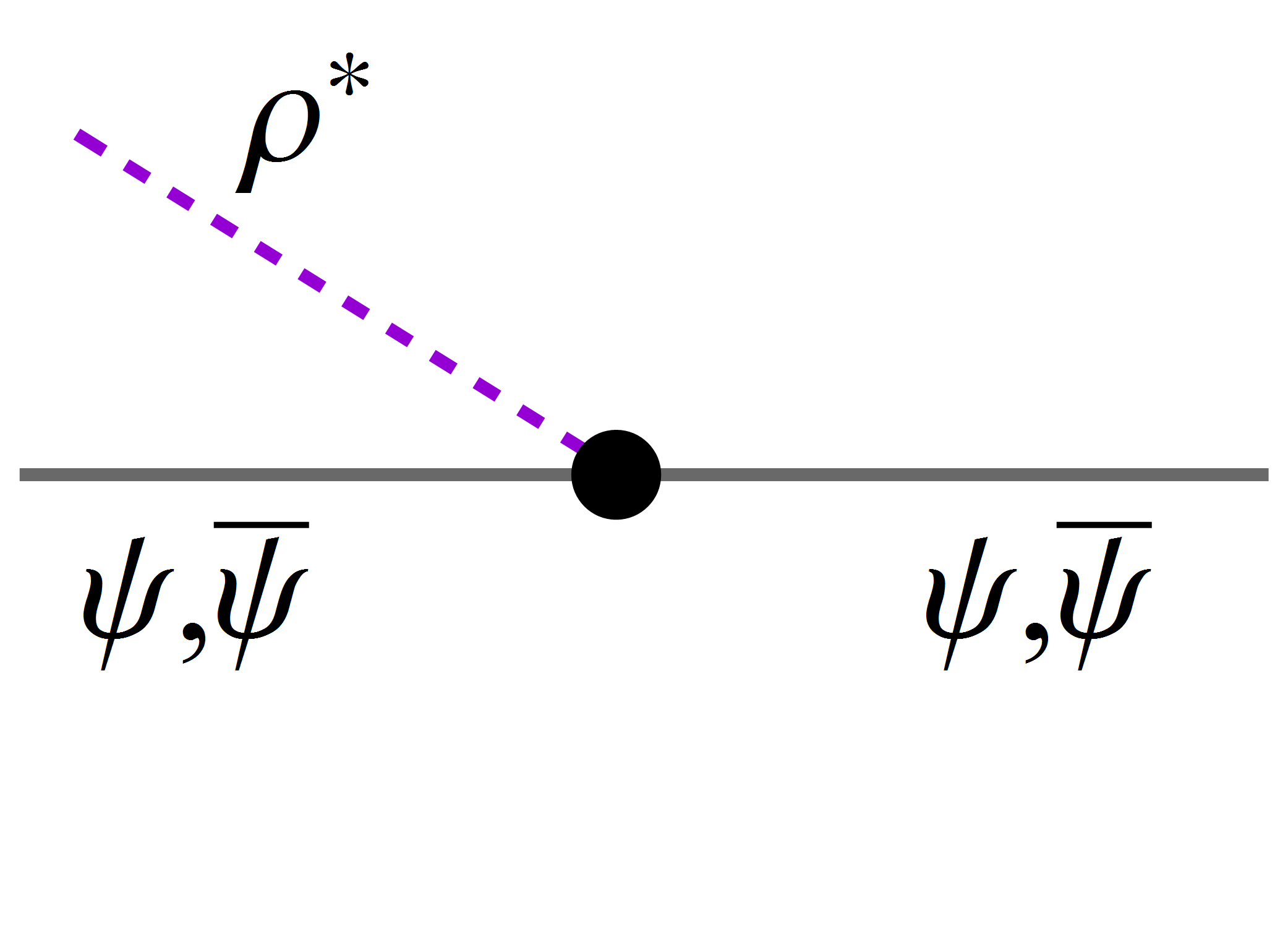}\hspace{15mm}
	\includegraphics[width=0.32\columnwidth]{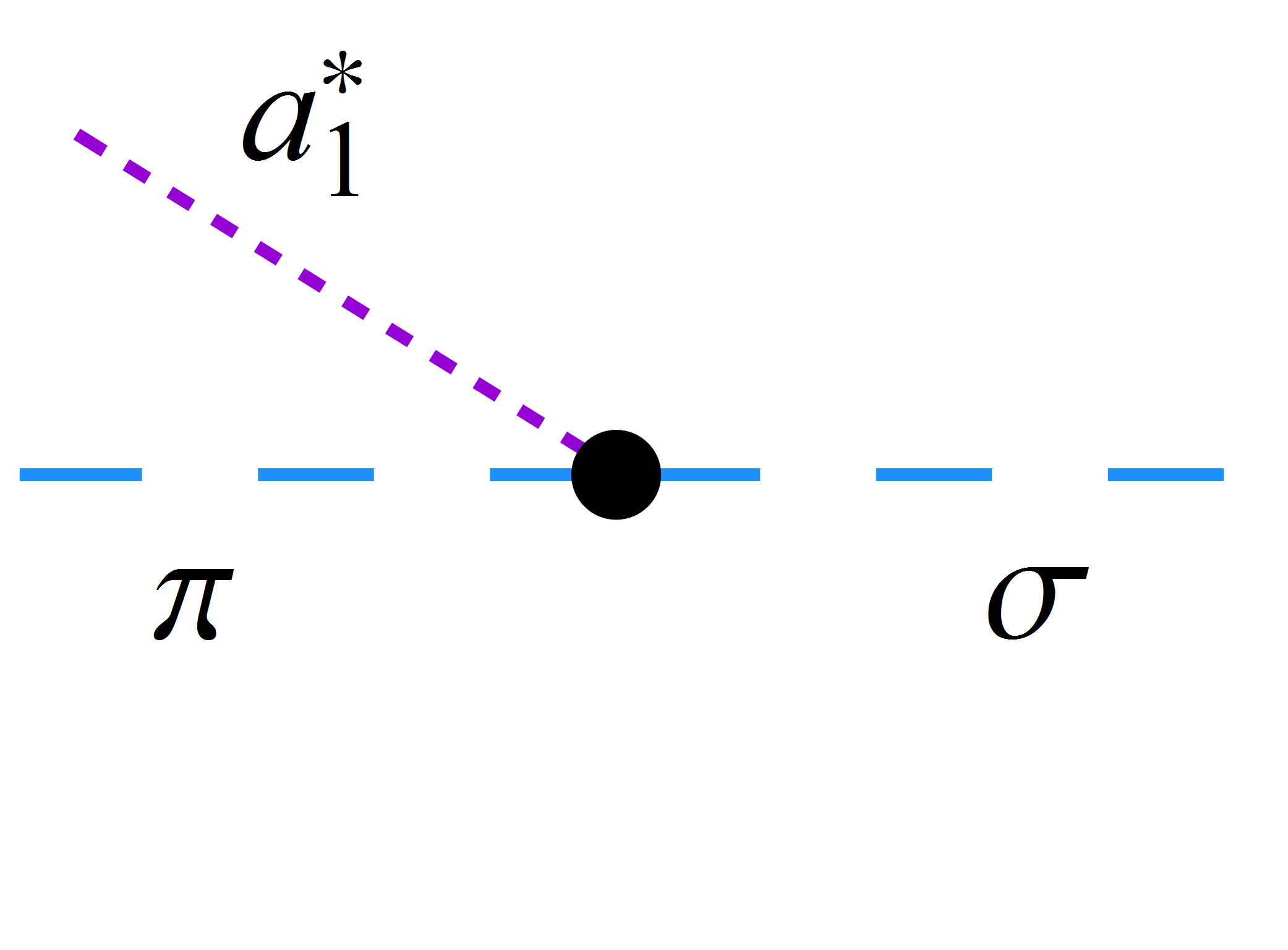}\\[3mm]
	\hspace{43mm}
	\includegraphics[width=0.32\columnwidth]{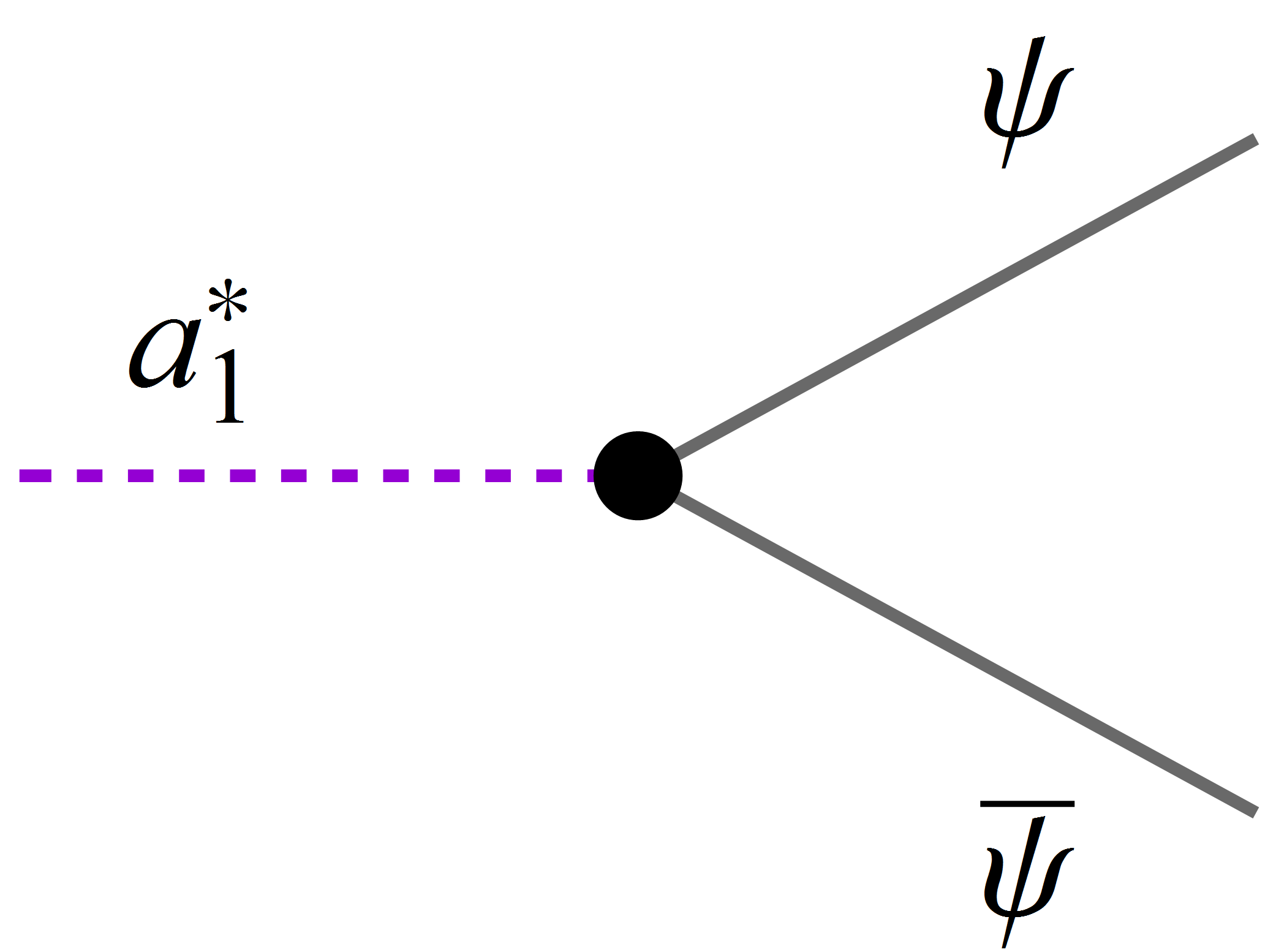}\hspace{15mm}
	\hspace{43mm}
	\includegraphics[width=0.32\columnwidth]{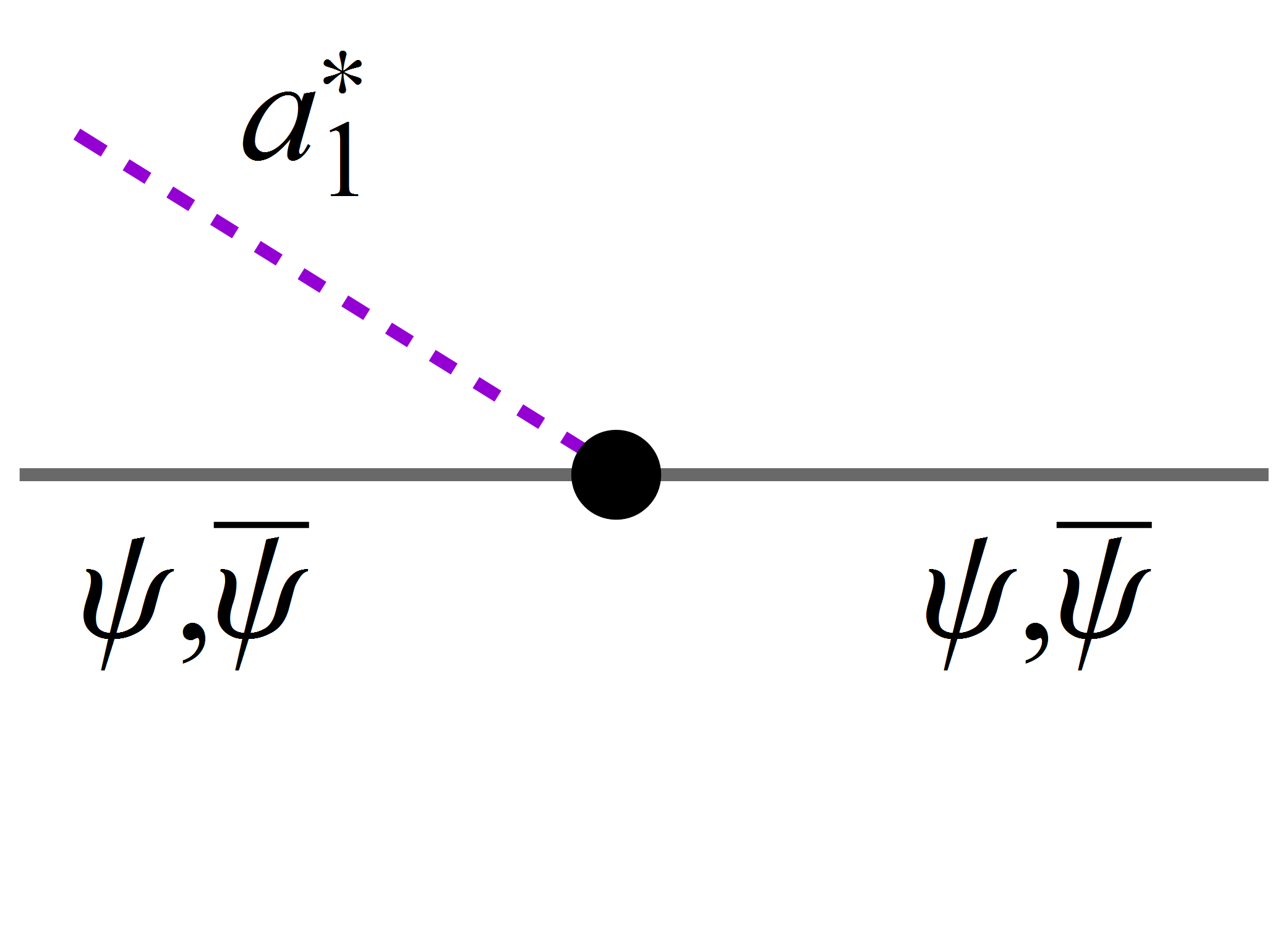}
	\caption{(color online) Collection of the possible time-like, $p^2=\omega^2-\vec{p}^{\,2}>0$, and space-like, $p^2<0$, processes. Asterisks denote off-shell particles with total energy $\omega$ and spatial momentum $\vec{p}$ while others represent on-shell particles from the heat bath. First column: time-like decay channels for the $\rho$ meson. Second column: time-like decay channels for the $a_1$ meson. Third column: absorption of a space-like $\rho$ excitation. Fourth column: absorption of a space-like $a_1$ excitation. In addition, particle-hole excitations on the Fermi surface of (anti-)quarks are possible, see text for details.}
	\label{fig:processes} 
\end{figure*}

In this section we summarize and discuss the different scattering and absorption or emission processes that can occur within our framework. We divide these processes into time-like and a space-like. Time-like processes involve off-shell particles with a total energy $\omega$ that is larger than their spatial momentum $|\vec{p}|$, i.e.~$\omega\geq |\vec{p}|$, while space-like processes involve a particle excitation with $\omega< |\vec{p}|$, see Fig.~\ref{fig:processes}.

The available time-like processes for an external off-shell $\rho$ meson, denoted as $\rho^*$, are given by
\begin{align}
	\label{eq:rho_pi_pi_decay}
	&\rho^*\rightarrow \pi + \pi\,, & \omega \geq \sqrt{(2m_\pi)^2+\vec{p}^{\,2}},\\
	\label{eq:rho_quark_quark_decay}
	&\rho^*\rightarrow \bar{\psi} + \psi\,,&  \omega \geq \sqrt{(2m_\psi)^2+\vec{p}^{\,2}},
\end{align}
where the kinematic constraints follow from energy conservation. We note that particles without asterisks represent on-shell particles and that their masses are given by the Euclidean masses in our truncation. Moreover, if there are particles available from the heat bath, the inverse of the above processes is also possible, giving rise to an equilibrium between direct and inverse processes. The corresponding statistical weight factors can be easily read off the corresponding loop functions, see the discussion in App.~\ref{sec:loop_functions}.

The time-like processes for an $a_1^*$ meson are given by
\begin{align}
	\label{eq:a1_sigma_pion_decay}
	&a_1^*\rightarrow \sigma + \pi\,, & \omega \geq \sqrt{(m_\sigma+m_\pi)^2+\vec{p}^{\,2}},\\
	\label{eq:a1_pion_sigma_capture}
	&a_1^*+\pi \rightarrow \sigma\,,& |\vec{p}|\leq \omega \leq (m_\sigma-m_\pi)\sqrt{1+\tfrac{\vec{p}^{\,2}}{\Delta m^2}},\\
	\label{eq:a1_sigma_pion_capture}
	&a_1^*+\sigma \rightarrow \pi\,,& |\vec{p}|\leq \omega \leq (m_\pi-m_\sigma)\sqrt{1+\tfrac{\vec{p}^{\,2}}{\Delta m^2}},\\
	\label{eq:a1_quark_quark_decay}
	&a_1^*\rightarrow \bar{\psi} + \psi\,,&  \omega \geq \sqrt{(2m_\psi)^2+\vec{p}^{\,2}},
\end{align}
with $\Delta m^2\equiv (m_\sigma-m_\pi)^2$. The first and last process describe standard decay channels while (\ref{eq:a1_pion_sigma_capture}) and (\ref{eq:a1_sigma_pion_capture}) describe capture processes which are only possible if there are particles from the heat bath available and if the kinematic constraints are fulfilled. In general, only the capture process $a_1^*+\pi \rightarrow \sigma$ is possible since the pion mass is usually smaller than the sigma mass.

We now turn to the space-like processes which, in the case of the $\rho$ meson given by
\begin{equation}
\label{eq:rho_spacelike}
\left.
\begin{aligned}
&\rho^* + \pi \rightarrow \pi\qquad\\
&\rho^* + \psi \rightarrow \psi\qquad
\end{aligned}
\right\}
\qquad 0\leq \omega \leq |\vec{p}|
\end{equation}
where $\psi$ represents both quarks and anti-quarks. These processes describe the absorption or, when considering the inverse processes, the emission of a space-like $\rho$ meson. 

Similarly, the space-like processes for the $a_1$ meson are given by
\begin{equation}
\label{eq:a1_spacelike}
\left.
\begin{aligned}
&a_1^* + \sigma \rightarrow \pi\qquad\\
&a_1^* + \pi \rightarrow \sigma\qquad\\
&a_1^* + \psi \rightarrow \psi\qquad
\end{aligned}
\right\}
\qquad 0\leq \omega \leq |\vec{p}|,
\end{equation}
with $\psi$ again representing both quarks and anti-quarks. We note that all space-like processes are only possible if there are particles from the heat bath available.

In addition to the processes discussed so far, also particle-hole excitations of quarks or anti-quarks are possible at finite chemical potential as soon as a Fermi sphere starts building up. These processes can be induced by an external $\rho$ or $a_1$ meson with an energy-momentum configuration that allows for an excitation of particle-hole pairs in the vicinity of the Fermi surface, see also App.~\ref{sec:loop_functions}. These particle-hole processes lead to interesting phenomena such as long-wavelength collective excitations such as sound waves which will be discussed in future applications of our approach.

\section{Explicit expressions for the loop functions}
\label{sec:loop_functions}
In this section we provide explicit expressions for the loop functions appearing in the flow equations for the retarded two-point functions, cf. App.~\ref{sec:flow_equations}.

The loop functions are written in a form that allows for an easy interpretation in terms of physical processes. We divide these processes into three categories: vacuum processes, capture processes, and particle-hole processes. Processes that can occur in the vacuum are associated to statistical weight factors of the form
\begin{align}
(1+ n_B(E_\alpha))(1+n_B(E_\beta))-n_B(E_\alpha)n_B(E_\beta), \nonumber
\end{align}
which relates to the process $\omega\rightarrow E_\alpha+E_\beta$ and its inverse process $E_\alpha+E_\beta\rightarrow \omega$. Capture processes are only possible when there are particles from a heat bath available. They carry weight factors like
\begin{align}
n_B(E_\alpha)(1+n_B(E_\beta))-n_B(E_\beta)(1+n_B(E_\alpha)) \nonumber
\end{align}
and correspond to processes like $\omega+E_\alpha\rightarrow E_\beta$ and its inverse. Particle-hole processes are proportional to the derivative of the occupation number, e.g.
$ n'_B(E_\alpha) $.

The momentum independent loop functions $I_{k,\rho/a_1}$ only contribute to the real part of the retarded two-point functions. They are given by
\begin{alignat}{2}
&I_{k,\rho}^{\pi} &&= \frac{k^4 g^2}{3 E_{k,\pi}^3\pi^2}\left[1+2n_B(E_{k,\pi})-2E_{k,\pi}n'_B(E_{k,\pi})\right],\nonumber \\
&I_{k,a_1}^{\alpha} &&= \frac{k^4 g^2}{6 E_{k,\alpha}^3\pi^2}\left[1+2n_B(E_{k,\alpha})-2E_{k,\alpha}n'_B(E_{k,\alpha})\right], \nonumber
\end{alignat}
with $\alpha\in\{\pi,\sigma\}$.

The momentum dependent loop functions $J_{k,\rho/a_1}(\omega)$ contribute to both, real and imaginary part, and therefore encode information about possible decay processes. Split into contributions related to the different processes they are listed in Eqs.~(\ref{eqn:JPionPion}-\ref{eqn:JQuarkQuark}). For the sake of simplicity we drop the external spatial momentum dependence here. In this case the theta function in the regulator functions, Eqs.~(\ref{eqs:regulator_scalar}-\ref{eqs:regulator_quarks}), has to be evaluated numerically. For the external energy $\omega$ we use the notation \mbox{$\omega_+ \rightarrow \omega + \mathrm{i}\epsilon$}.

The loop function $J_{k,\rho}^{\pi\pi}(\omega)$, Eq.~(\ref{eqn:JPionPion}), describes the decay of an off-shell $\rho$ meson  into a pion pair, \mbox{$\rho^*\rightarrow \pi+\pi$}, and its inverse process \mbox{$\pi+\pi\rightarrow \rho^*$}. The imaginary part of these processes is non-vanishing when it becomes energetically possible, e.g. $\omega\geq 2E_{k,\pi}$ for the former process. The mesonic decay channel for the $a_1$ meson is described by the loop function $J_{k,a_1}^{\alpha\beta}(\omega)$, Eq.~(\ref{eqn:JAlphaBeta}), with $\alpha,\beta\in \{\pi,\sigma\}$ but distinguished. Here we also have the vacuum processes \mbox{$a_1^*\rightarrow \alpha+\beta$} and \mbox{$\alpha+\beta\rightarrow a_1^*$} but also capture processes as described above. These processes are of the form \mbox{$a_1^*+\alpha/\beta\rightarrow \beta/\alpha$} and are described by single occupation numbers $\pm n_B(E_{k,\alpha/\beta})$. For $T\rightarrow 0$ these contributions vanish completely, in contrast to the vacuum processes. The mesonic decay in a quark-antiquark pair is described by the loop function $J_{k,\alpha}^{\psi\bar{\psi}}(\omega)$, Eq.~(\ref{eqn:JQuarkQuark}). Here we also have the vacuum process \mbox{$\rho^*/a_1^*\rightarrow \psi+\bar{\psi}$} and its inverse \mbox{$\psi+\bar{\psi}\rightarrow \rho^*/a_1^*$}. The form \mbox{$(1-n_F(E_{k,\psi}))$} reflects Pauli blocking: available states for the decay products are suppressed since there are real fermions in the heat bath. The coefficients
$\tilde{E}_{1,k}^{(\alpha)},\tilde{E}_{2,k}^{(\alpha)}$ and $\tilde{E}_{3,k}^{(\alpha)}$ are defined by 
\begin{alignat}{2}
&\tilde{E}_{1,k}^{(\alpha)} &&= 8 k^2+L_k^{(\alpha)},\\
&\tilde{E}_{2,k}^{(\alpha)} &&= 8m_{k,\psi}^2-L_k^{(\alpha)},\\
&\tilde{E}_{3,k}^{(\alpha)} &&= -4E_{k,\psi}^2-8m_{k,\psi}^2+L_k^{(\alpha)},
\end{alignat}
with $L_k^{(\rho)}=12m_{k,\psi}^2$ and $L_k^{(a_1)}=0$. 

Additionally, there are terms proportional to a derivative of an occupation number which are connected to particle-hole processes.

\begin{widetext}
\begin{alignat}{3}
J_{k,\rho}^{\pi\pi}(\omega, |\vec{p}|=0) = 
&\quad-\frac{1}{(\omega_+ -2E_{k,\pi})}
&&\frac{k^6 g^2 m_{k,\rho}^4}{15 \pi^2 E_{k,\pi}^3m_{k,a_1}^4}
&&\quad\left[1+2n_B(E_{k,\pi})-E_{k,\pi}n'_B(E_{k,\pi})\right]\nonumber\\
&\quad+\frac{1}{(\omega_+ -2E_{k,\pi})^2}
&&\frac{k^6 g^2 m_{k,\rho}^4}{15 \pi^2 E_{k,\pi}^3m_{k,a_1}^4}
&&\quad\left[1+2n_B(E_{k,\pi})\right]\nonumber\\
&\quad+\frac{1}{(\omega_+ +2E_{k,\pi})}
&&\frac{k^6 g^2 m_{k,\rho}^4}{15 \pi^2 E_{k,\pi}^3m_{k,a_1}^4}
&&\quad\left[1+2n_B(E_{k,\pi})-E_{k,\pi}n'_B(E_{k,\pi})\right]\nonumber\\
&\quad+\frac{1}{(\omega_+ +2E_{k,\pi})^2}
&&\frac{k^6 g^2 m_{k,\rho}^4}{15 \pi^2 E_{k,\pi}^3m_{k,a_1}^4}
&&\quad\left[1+2n_B(E_{k,\pi})\right]\label{eqn:JPionPion}
\end{alignat}
\begin{alignat}{3}
J_{k,a_1}^{\alpha\beta}(\omega, |\vec{p}|=0) = 
&\quad+\frac{1}{(\omega_+ +E_{k,\alpha}+E_{k,\beta})}
&&\frac{k^6g^2}{30\pi^2 E_{k,\alpha}^3E_{k,\beta}}
&&\quad\left[1+n_B(E_{k,\alpha})+n_B(E_{k,\beta})-E_{k,\alpha}n'_B(E_{k,\alpha})\right]\nonumber\\
&\quad+\frac{1}{(\omega_+ +E_{k,\alpha}+E_{k,\beta})^2}
&&\frac{k^6g^2}{30\pi^2 E_{k,\alpha}^2E_{k,\beta}}
&&\quad\left[1+n_B(E_{k,\alpha})+n_B(E_{k,\beta})\right]\nonumber\\
&\quad-\frac{1}{(\omega_+ -E_{k,\alpha}-E_{k,\beta})}
&&\frac{k^6g^2}{30\pi^2 E_{k,\alpha}^3E_{k,\beta}}
&&\quad\left[1+n_B(E_{k,\alpha})+n_B(E_{k,\beta})-E_{k,\alpha}n'_B(E_{k,\alpha})\right]\nonumber\\
&\quad+\frac{1}{(\omega_+ -E_{k,\alpha}-E_{k,\beta})^2}
&&\frac{k^6g^2}{30\pi^2 E_{k,\alpha}^2E_{k,\beta}}
&&\quad\left[1+n_B(E_{k,\alpha})+n_B(E_{k,\beta})\right]\nonumber\\
&\quad+\frac{1}{(\omega_+ -E_{k,\alpha}+E_{k,\beta})}
&&\frac{k^6g^2}{30\pi^2 E_{k,\alpha}^3E_{k,\beta}}
&&\quad\left[n_B(E_{k,\alpha})-n_B(E_{k,\beta})-E_{k,\alpha}n'_B(E_{k,\alpha})\right]\nonumber\\
&\quad+\frac{1}{(\omega_+ -E_{k,\alpha}+E_{k,\beta})^2}
&&\frac{k^6g^2}{30\pi^2 E_{k,\alpha}^2E_{k,\beta}}
&&\quad\left[-n_B(E_{k,\alpha})+n_B(E_{k,\beta})\right]\nonumber\\
&\quad+\frac{1}{(\omega_+ +E_{k,\alpha}-E_{k,\beta})}
&&\frac{k^6g^2}{30\pi^2 E_{k,\alpha}^3E_{k,\beta}}
&&\quad\left[-n_B(E_{k,\alpha})+n_B(E_{k,\beta})+E_{k,\alpha}n'_B(E_{k,\alpha})\right]\nonumber\\
&\quad+\frac{1}{(\omega_+ +E_{k,\alpha}-E_{k,\beta})^2}
&&\frac{k^6g^2}{30\pi^2 E_{k,\alpha}^2E_{k,\beta}}
&&\quad\left[-n_B(E_{k,\alpha})+n_B(E_{k,\beta})\right]\label{eqn:JAlphaBeta}
\end{alignat}
\begin{alignat}{3}
J_{k,\alpha}^{\psi\bar{\psi}}(\omega, |\vec{p}|=0) = 
&\quad - \frac{1}{(\omega_+)}
&&\frac{k^4h_V^2}{6\pi^2 E_{k,\psi}^3}
&&\quad\left[\tilde{E}_{3,k}^{(\alpha)}(n'_F(E_{k,\psi}+\mu)-n'_F(E_{k,\psi}-\mu))\right]\nonumber\\
&\quad -\frac{1}{(\omega_+-2E_{k,\psi})}
&&\frac{k^4h_V^2}{6\pi^2 E_{k,\psi}^4}
&&\quad\left[\tilde{E}_{2,k}^{(\alpha)}(1-n_F(E_{k,\psi}-\mu)-n_F(E_{k,\psi}+\mu))-E_{k,\psi}\tilde{E}_{1,k}^{(\alpha)}n'_F(E_{k,\psi}+\mu)\right]\nonumber\\
&\quad -\frac{1}{(\omega_+-2E_{k,\psi})^2}
&&\frac{k^4h_V^2 }{6\pi^2 E_{k,\psi}^3}
&&\quad\left[\tilde{E}_{1,k}^{(\alpha)}(1-n_F(E_{k,\psi}-\mu)-n_F(E_{k,\psi}+\mu))\right]\nonumber\\
&\quad +\frac{1}{(\omega_++2E_{k,\psi})}
&&\frac{k^4h_V^2}{6\pi^2 E_{k,\psi}^4}
&&\quad\left[\tilde{E}_{2,k}^{(\alpha)}(1-n_F(E_{k,\psi}-\mu)-n_F(E_{k,\psi}+\mu))-E_{k,\psi}\tilde{E}_{1,k}^{(\alpha)}n'_F(E_{k,\psi}-\mu)\right]\nonumber\\
&\quad -\frac{1}{(\omega_++2E_{k,\psi})^2}
&&\frac{k^4h_V^2}{6\pi^2 E_{k,\psi}^3}
&&\quad\left[\tilde{E}_{1,k}^{(\alpha)}(1-n_F(E_{k,\psi}-\mu)-n_F(E_{k,\psi}+\mu))\right]\label{eqn:JQuarkQuark}
\end{alignat}
\end{widetext}
\bibliography{QCD}

\end{document}